\documentclass[12pt]{article}
\pdfoutput=1
\usepackage[utf8x]{inputenc}
\usepackage[colorlinks,linkcolor=blue,citecolor=blue,bookmarks,bookmarksnumbered]{hyperref}
\usepackage[scaled=0.85]{helvet}
\usepackage{amsmath,amssymb,accents,mathrsfs,XoohmE}
\usepackage{graphicx,color}
\usepackage{booktabs}
\usepackage{multirow}
\usepackage{placeins}
\usepackage{amsmath}

\usepackage{cite}

\usepackage{enumitem}

\usepackage{XoohmE}

\def\rD{{\rm D}}

\def\bDb{\hbox{\kern2pt\vrule height10pt depth-9.2pt width6pt\kern-9pt{$\boldsymbol D$}}\mkern-2mu}

\def\bQb{\hbox{\kern2pt\vrule height10pt depth-9.2pt width6pt\kern-9pt{$\boldsymbol Q$}}}

\def\BF{{\bs\Phi}}
\def\bBF{{\Bar{\bs{\Phi}}}}

\def\BS{{\bs\Sigma}}
\def\bBS{{\Bar{\bs{\Sigma}}}}

\def\a{\alpha}
\def\b{\beta}
\def\c{\chi}
\def\d{\delta}

\def\g{\gamma}

\def\i{\iota}
\def\j{\psi}
\def\k{\kappa}
\def\l{\lambda}

\def\o{\omega}

\def\r{\rho}
\def\s{\sigma}
\def\t{\tau}

\def\z{\zeta}

\def\L{\Lambda}
\def\O{\Omega}
\def\P{\Pi}

\def\S{\Sigma}

\def\ca{{\cal A}}
\def\cb{{\cal B}}
\def\cc{{\cal C}}
\def\cd{{\cal D}}
\def\ce{{\cal E}}
\def\cf{{\cal F}}

\def\cj{{\cal J}}

\def\cl{{\cal L}}

\def\cn{{\cal N}}

\def\cp{{\cal P}}
\def\cq{{\cal Q}}

\def\cy{{\cal Y}}

\def\tca{\Tilde{\cal A}}
\def\tcb{\Tilde{\cal B}}
\def\tcc{\Tilde{\cal C}}
\def\tcd{\Tilde{\cal D}}
\def\vca{\Check{\cal A}}
\def\vcb{\Check{\cal B}}
\def\vcc{\Check{\cal C}}
\def\vcd{\Check{\cal D}}
\def\vce{\Check{\cal E}}
\def\hca{\Hat{\cal A}}
\def\hcb{\Hat{\cal B}}
\def\hcc{\Hat{\cal C}}
\def\hcd{\Hat{\cal D}}
\def\hce{\Hat{\cal E}}
\def\hcf{\Hat{\cal F}}
\def\hcg{\Hat{\cal G}}

\definecolor{Green}  {rgb}{0.10,0.70,0.10} 
\definecolor{Orange} {rgb}{1.00,0.50,0.15} 
\definecolor{Red}    {rgb}{0.90,0.00,0.12} 
\definecolor{Purple} {rgb}{0.50,0.25,0.55} 
\definecolor{Turque} {rgb}{0.00,0.65,0.85} 
\definecolor{Blue}   {rgb}{0.00,0.00,1.00} 
\definecolor{Magenta}{rgb}{1.00,0.00,1.00} 
\definecolor{Gold}   {rgb}{1.00,0.75,0.25} 
\definecolor{Seaweed}{rgb}{0.01,0.24,0.09} 
\definecolor{Brown}  {rgb}{0.43,0.26,0.32} 
\definecolor{grey1}  {rgb}{0.20,0.20,0.20} 
\definecolor{grey2}  {rgb}{0.40,0.40,0.40} 
\definecolor{grey3}  {rgb}{0.60,0.60,0.60} 
\definecolor{grey4}  {rgb}{0.80,0.80,0.80} 
\definecolor{grey5}  {rgb}{0.90,0.90,0.90} 
\def\C#1#2{{\ifcase#1\or
             \color{Green}\or \color{Orange}\or \color{Red}\or
              \color{Purple}\or \color{Turque}\or \color{Blue}\or
               \color{Magenta}\or \color{Gold}\or \color{Seaweed}\or
                \color{Brown}\or\color{grey1}\or\color{grey2}\or
                 \color{grey3}\else\color{grey4}\fi#2}}

\definecolor{Slate} {rgb}{0.00,0.45,0.55}

\usepackage[enableskew,vcentermath]{youngtab}

\definecolor{Hey}{rgb}{.9,.05,.4}
\definecolor{orange}{rgb}{1,.5,0}
\definecolor{plum}{rgb}{.4,0,.6}
\definecolor{R}{rgb}{1,0,0}
\definecolor{G}{rgb}{0.1,0.7,0}
\definecolor{B}{rgb}{0,0,1}
\usepackage{lipsum}
\usepackage{listings}
\definecolor{MyDarkGreen}{rgb}{0.0,0.4,0.0} 
\def\fracm#1#2{\hbox{\large{${\frac{{#1}}{{#2}}}$}}}

\def\be{\begin{equation}}
\def\ee{\end{equation}}
\newcommand{\bea}{\begin{eqnarray}}
\newcommand{\eea}{\end{eqnarray}}
\newcommand{\ena}{\end{eqnarray}}

\def\pp{{\mathchoice
          {
              \kern 1pt%
              \raise 1pt
              \vbox{\hrule width5pt height0.4pt depth0pt
                    \kern -2pt
                    \hbox{\kern 2.3pt
                          \vrule width0.4pt height6pt depth0pt
                          }
                    \kern -2pt
                    \hrule width5pt height0.4pt depth0pt}%
                    \kern 1pt
           }
            {
              \kern 1pt%
              \raise 1pt
              \vbox{\hrule width4.3pt height0.4pt depth0pt
                    \kern -1.8pt
                    \hbox{\kern 1.95pt
                          \vrule width0.4pt height5.4pt depth0pt
                          }
                    \kern -1.8pt
                    \hrule width4.3pt height0.4pt depth0pt}%
                    \kern 1pt
            }
            {
              \kern 0.5pt%
              \raise 1pt
              \vbox{\hrule width4.0pt height0.3pt depth0pt
                    \kern -1.9pt  
                    \hbox{\kern 1.85pt
                          \vrule width0.3pt height5.7pt depth0pt
                          }
                    \kern -1.9pt
                    \hrule width4.0pt height0.3pt depth0pt}%
                    \kern 0.5pt
            }
            {
              \kern 0.5pt%
              \raise 1pt
              \vbox{\hrule width3.6pt height0.3pt depth0pt
                    \kern -1.5pt
                    \hbox{\kern 1.65pt
                          \vrule width0.3pt height4.5pt depth0pt
                          }
                    \kern -1.5pt
                    \hrule width3.6pt height0.3pt depth0pt}%
                    \kern 0.5pt
            }
        }}

\def\mm{{\mathchoice
                       {
                             \kern 1pt
               \raise 1pt    \vbox{\hrule width5pt height0.4pt depth0pt
                                  \kern 2pt
                                  \hrule width5pt height0.4pt depth0pt}
                             \kern 1pt}
                       {
                            \kern 1pt
               \raise 1pt \vbox{\hrule width4.3pt height0.4pt depth0pt
                                  \kern 1.8pt
                                  \hrule width4.3pt height0.4pt depth0pt}
                             \kern 1pt}
                       {
                            \kern 0.5pt
               \raise 1pt
                            \vbox{\hrule width4.0pt height0.3pt depth0pt
                                  \kern 1.9pt
                                  \hrule width4.0pt height0.3pt depth0pt}
                            \kern 1pt}
                       {
                           \kern 0.5pt
             \raise 1pt  \vbox{\hrule width3.6pt height0.3pt depth0pt
                                  \kern 1.5pt
                                  \hrule width3.6pt height0.3pt depth0pt}
                           \kern 0.5pt}
                       }}

\def\ad{{\kern0.5pt
                   \alpha \kern-5.05pt \raise5.8pt\hbox{$\textstyle.$}\kern
0.5pt}}

\def\qd{{\kern0.5pt
                   q \kern-5.05pt \raise5.8pt\hbox{$\textstyle.$}\kern
0.5pt}}
\def\Dot#1{{\kern0.5pt
     {#1} \kern-5.05pt \raise5.8pt\hbox{$\textstyle.$}\kern
0.5pt}}

\catcode`@=11
\def\un#1{\relax\ifmmode\@@underline#1\else
        $\@@underline{\hbox{#1}}$\relax\fi}
\catcode`@=12

\def\dslash{\not{\hbox{\kern-2pt $\partial$}}}
\def\Dslash{\not{\hbox{\kern-4pt $D$}}}
\def\pslash{\not{\hbox{\kern-2.3pt $p$}}}
 \newtoks\slashfraction
 \slashfraction={.13}
 \def\slash#1{\setbox0\hbox{$ #1 $}
 \setbox0\hbox to \the\slashfraction\wd0{\hss \box0}/\box0 }
 
\def\kcr{{\hbox{\ro \char'170}}}                
\def\ktl{{\hbox{\ro \char'170}}}        
\def\ktr{{\hbox{\ro \char'170}}}        
\def\kbl{{\hbox{\ro \char'170}}}        
\def\kbr{{\hbox{\ro \char'170}}}        

\def\plpl{\raise-2pt\hbox{$\raise3pt\hbox{$_+$}\hskip-6.67pt\raise0.0pt
\hbox{$^+$}\hskip 0.01pt$}}
\def\mimi{\raise-2pt\hbox{$\raise3pt\hbox{$_-$}\hskip-6.67pt\raise0.0pt
\hbox{$^-$}\hskip 0.01pt$}}

\def\bo{{\raise.15ex\hbox{\large$\Box$}}}               
\def\pa{\partial}                                       
\def\TH{{\raise.2ex\hbox{$\displaystyle \bigodot$}\mskip-4.7mu \llap H \;}}
\def\face{{\raise.2ex\hbox{$\displaystyle \bigodot$}\mskip-2.2mu \llap {$\ddot
        \smile$}}}                                      

\def\dt#1{\on{\hbox{\bf .}}{#1}}                
\def\Dot#1{\dt{#1}}

\def\Tilde#1{\widetilde{#1}}                    
\def\Hat#1{\widehat{#1}}                        
\def\Bar#1{\overline{#1}}                       
\def\leftrightarrowfill{$\mathsurround=0pt \mathord\leftarrow \mkern-6mu
        \cleaders\hbox{$\mkern-2mu \mathord- \mkern-2mu$}\hfill
        \mkern-6mu \mathord\rightarrow$}
\def\dvec#1{\vbox{\ialign{##\crcr
        \leftrightarrowfill\crcr\noalign{\kern-1pt\nointerlineskip}
        $\hfil\displaystyle{#1}\hfil$\crcr}}}           
\def\dt#1{{\buildrel {\hbox{\LARGE .}} \over {#1}}}     

\def\fracm#1#2{\hbox{\large{${\frac{{#1}}{{#2}}}$}}}
\def\sfrac#1#2{{\vphantom1\smash{\lower.5ex\hbox{\small$#1$}}\over
        \vphantom1\smash{\raise.4ex\hbox{\small$#2$}}}} 
\def\bfrac#1#2{{\vphantom1\smash{\lower.5ex\hbox{$#1$}}\over
        \vphantom1\smash{\raise.3ex\hbox{$#2$}}}}       
\def\afrac#1#2{{\vphantom1\smash{\lower.5ex\hbox{$#1$}}\over#2}}    

\def\pa{\partial}      
\let\bm\relax
\newcommand{\bm}[1]{{\boldsymbol{#1}}}

\def\ad{{\Dot{\alpha}}}

 \font\rOpe=cmsy10                        
 \def\ktl{{\hbox{\rOpe\char'170}}}        
 \def\kbl{{\hbox{\rOpe\char'170}}}        
 \def\kcr{{\reflectbox{\rOpe\char'170}}}        
 \def\ktr{{\reflectbox{\rOpe\char'170}}}        
 \def\kbr{{\reflectbox{\rOpe\char'170}}}        
 \def\Border{\vbox{\hsize0pt
        \setlength{\unitlength}{1mm}
        \newcount\xco
        \newcount\yco
        \xco=-21
        \yco=12
        \begin{picture}(0,0)(-7.5,0)
        \put(\xco,\yco){$\ktl$}
        \advance\yco by-1
        {\loop
        \put(\xco,\yco){$\kcr$}
        \advance\yco by-2
        \ifnum\yco>-240
        \repeat
        \put(\xco,\yco){$\kbl$}}
        \xco=170
        \yco=12
        \put(\xco,\yco){$\ktr$}
        \advance\yco by-1
        {\loop
        \put(\xco,\yco){$\kcr$}
        \advance\yco by-2
        \ifnum\yco>-240
        \repeat
        \put(\xco,\yco){$\kbr$}}
        \put(-19.5,13){\scalebox{.6065}{%
         University of Maryland Center for String and Particle  Theory \&\ Physics Department%
        |University of Maryland Center for String and Particle  Theory \&\ Physics Department}}
        \put(-19.5,-241.5){\scalebox{.5835}{%
         ****University of Maryland * Center for String and
         Particle  Theory* Physics Department****University of Maryland *Center
        for String and Particle  Theory* Physics Department}}
        \end{picture}
        \par\vskip-8mm}}
\definecolor{UMred}{rgb}{.9,.05,.2}
\definecolor{HUblue}{rgb}{.0,.3,.7}

\definecolor{Red}    {rgb}{0.90,0.00,0.12} 
\definecolor{Blue}   {rgb}{0.00,0.00,1.00} 
\definecolor{Green}  {rgb}{0.10,0.70,0.10} 
\definecolor{Turque} {rgb}{0.00,0.65,0.85} 
\definecolor{Orange} {rgb}{1.00,0.50,0.15} 
\definecolor{Magenta}{rgb}{1.00,0.00,1.00} 
\definecolor{Gold}   {rgb}{1.00,0.75,0.25} 
\definecolor{Seaweed}{rgb}{0.01,0.24,0.09} 
\definecolor{Purple} {rgb}{0.50,0.25,0.55} 
\definecolor{Brown}  {rgb}{0.43,0.26,0.32} 
\definecolor{grey1}  {rgb}{0.20,0.20,0.20} 
\definecolor{grey2}  {rgb}{0.40,0.40,0.40} 
\definecolor{grey3}  {rgb}{0.60,0.60,0.60} 
\definecolor{grey4}  {rgb}{0.80,0.80,0.80} 
\definecolor{grey5}  {rgb}{0.90,0.90,0.90} 
\def\C#1#2{{\ifcase#1\or
             \color{Red}\or \color{Green}\or \color{Blue}\or\
              \color{Turque}\or \color{Orange}\or \color{Magenta}\or
               \color{Gold}\or \color{Seaweed}\or \color{Purple}\or
                \color{Brown}\or\color{grey1}\or\color{grey2}\or
                 \color{grey3}\else\color{grey4}\fi#2}}

\definecolor{Slate} {rgb}{0.00,0.45,0.55}

\newdimen\parshift\parshift=\parindent
\catcode`@=11
 \long\def\@footnotetext#1{\insert\footins{\reset@font\footnotesize
           \interlinepenalty\interfootnotelinepenalty\splittopskip%
            \footnotesep\splitmaxdepth\dp\strutbox\floatingpenalty\@MM%
             \hsize\columnwidth\addtolength{\hsize}{-2\parindent}
              \@parboxrestore\protected@edef\@currentlabel%
              {\csname p@footnote\endcsname\@thefnmark}%
                \color@begingroup%
                 \@makefntext{\rule\z@\footnotesep\ignorespaces#1%
                  \@finalstrut\strutbox}%
                \color@endgroup}}
 \long\def\@makefntext#1{\hglue\parshift%
           \vbox{\noindent\baselineskip=11pt plus.5pt minus.5pt\hb@xt@0em{\hss\@makefnmark\kern1pt}#1}}
\catcode`@=12

\newskip\humongous \humongous=0pt plus 1000pt minus 1000pt

\newif\ifdtup

\makeatletter
\def\section{\@startsection{section}{1}{\z@}
        {3ex plus-1ex minus-.2ex}{1pt plus1pt}{\large\sf\bfseries\boldmath}}
\def\subsection{\@startsection{subsection}{2}{\z@}
         {1.5ex plus-1ex minus-.2ex}{0.01pt plus1pt}{\sf\slshape}}
\def\subsubsection{\@startsection{subsubsection}{3}{\z@}
          {1.5ex plus-1ex minus-.2ex}{0.01pt plus0.2pt}{\sf\boldmath}}
\def\paragraph{\@startsection{paragraph}{4}{\z@}
           {.75ex \@plus.5ex \@minus.2ex}{-2mm}{\sf\bfseries\boldmath}}
\makeatother

 \allowdisplaybreaks
 \seceq

\begin{document}

\thispagestyle{empty}
\noindent{\small
\hfill{$~~$}  \\ 
{}
}
\begin{center}
{\large \bf
On 1D, $\bm {\cal N}$ = 4 Supersymmetric SYK-Type Models (II)  
}   \\   [8mm]
{\large {
S.\ James Gates, Jr.\footnote{sylvester$_-$gates@brown.edu}${}^{,a, b}$,
Yangrui Hu\footnote{yangrui$_-$hu@brown.edu}${}^{,a,b}$, and
S.-N. Hazel Mak\footnote{sze$_-$ning$_-$mak@brown.edu}${}^{,a,b}$
}}
\\*[6mm]
\emph{
\centering
$^{a}$Brown Theoretical Physics Center,
\\[1pt]
Box S, 340 Brook Street, Barus Hall,
Providence, RI 02912, USA
\\[10pt]
and
\\[10pt]
$^{b}$Department of Physics, Brown University,
\\[1pt]
Box 1843, 182 Hope Street, Barus \& Holley,
Providence, RI 02912, USA
}
 \\*[80mm]
{ ABSTRACT}\\[05mm]
\parbox{142mm}{\parindent=2pc\indent\baselineskip=14pt plus1pt
This paper is an extension of our last 1D, $\cn=4$ supersymmetric SYK paper [arXiv:2103.11899]. In this paper we introduced the complex linear supermultiplet (CLS), which is ``usefully inequivalent'' to the chiral supermultiplet. 
We construct three types of models based on the complex linear supermultiplet containing quartic interactions from modified CLS kinetic term, quartic interactions from 3-pt vertices integrated over the whole superspace, and $2(q-1)$-pt interactions generated via superpotentials respectively. 
A strong evidence for the inevitability of dynamical bosons for 1D, $\cn=4$ SYK is also presented.
} \end{center}
\vfill
\vfill
\clearpage

\newpage
{\hypersetup{linkcolor=black}
\tableofcontents
}

\newpage
\section{Introduction}
\label{sec:NTRO}

Previously \cite{GHM}, we have studied the question of
constructing 1D, $\cn = 4$ extensions of SYK models \cite{SY,K1,K2} by using a technique of starting with 4D, $\cn = 1$
supermultiplets and compactifying them to 1D, $\cn = 4$
supermultiplets.  In 4D superspace, the chiral \cite{GL1,GL2,WZ,Wss,Fy8}, vector
\cite{VSM1,VSM2}, and tensor \cite{C-GuLL} supermultiplets are all valid candidates to take as starting
points.  The reason for this is the fact, that under such a
reduction, the notion of spin vanishes and one is simply left with a number of distinct 1D, $\cn = 4$ supermultiplets.
However, these are not the complete ``roll call of the 
1D, $\cn = 4$ supermultiplet zoo.'' There is one more
member, the 4D, $\cn = 1$ complex linear supermultiplet 
and it can be used as a starting point.  So
a more complete analysis requires including this final
supermultiplet and assessing its utility in the attempt
to construct the most general 1D, $\cn = 4$ supersymmetrical
extension of SYK-type models.

To accomplish this goal, it will be necessary to reexamine a little recognized property
first enunciated in a paper \cite{Neqv} with the title, ``Linear and chiral superfields 
are usefully inequivalent.''  This property asserts that a given on-shell spectrum of 
free component fields can have multiple embeddings within distinct superfields such that the most 
general supersymmetric interactions of these fields requires use of the multiple embeddings.  
The use of this principle, to our knowledge, first appeared in a 1984 publication \cite{TwsT} 
that uncovered ``twisted chiral superfields'' and ``twisted superpotentials'' before
their use in the important discovery of mirror symmetry.

The title of the paper by H\" ubsch \cite{Neqv} exactly describes the gateway for our
investigation of 4D, $\cal N$ = 1 actions that possess the required higher $n$-point
functions of fermions in the context where the spin spectrum of fields does not exceed
one-half.  Examples of such non-supersymmetrical Lagrangians \cite{ThRR,NmBJL1,NmBJL2,GN}, 
either regarded as effective actions or descriptions of fundamental physics, have a 
storied history in the field.  Perhaps the Fermi Universal Theory of the Weak 
Interactions is the greatest in this approach as it pointed the direction to 
the triumphal construction of the renormalizable QFT needed for the flavor interactions 
of the Standard Model.



In addition to chiral, vector and tensor supermultiplets, in this paper we also utilize the complex linear supermultiplet,
whose existence was introduced in the literature via a study \cite{SFSG} of superfield supergravity.
Thereafter, \cite{GS,SUSYBk,CNM} a literature began to evolve regarding this supermultiplet.
Like the chiral supermultiplet, its propagating degrees of freedom consist only of fields
with spins of one-half and zero.  However, unlike the real linear supermultiplet, there
are no gauge fields in its spectrum.  Thus, it is not a cavil against the chiral supermultiplet
to note that if one is interested in a 4D, $\cal N$ = 1 supersymmetrical theory that
possesses: \newline \indent
(a.) {\it {no}} gauge fields in it spectrum, \newline \indent
(b.) SYK type 2$n$-point couplings among fermions, and  \newline \indent
(c.) {\it {no}} higher derivative for the fermions, \newline \noindent
requires the use of the chiral supermultiplet {\it {together}} with the complex linear supermultiplet.

For the convenience of the reader, references \cite{SFSG,GS,SUSYBk,CNM,B&K,K1ps,CNM1,CNM2,CNM3,K4,CNM4,CNM4a,K5,K6,K7,K8,K9}
are provided in the bibliography of this work as a guide to the literature surrounding the development and study of the complex 
linear supermultiplet.  A bottom line demonstration accomplished in this work is
a proof of how essential and critical the useful inequivalence is to the construction of SYK 
type models with degrees of SUSY extension greater than two.  Our successful construction 
of models with $\cal N$ = 4 SUSY most certainly raises possibility that there may exist an 
ultimate such model at some maximal even larger value of $\cal N$ with unusual properties.

In the realm of phenomenological models that incorporate SUSY, it is essentially a 
universal truth that these are built upon the chiral supermultiplet.  With complex 
linear supermultiplet being all but ignored.  This implies there is an entire realm 
in the model space of such constructions that has never been explored.


We organize our paper in the following manner. In section \ref{sec:review}, we review the two-component conventions for discussing the chiral supermultiplet (CS), complex linear supermultiplet (CLS), vector supermultiplet (VS), and tensor supermultiplet (TS). Section \ref{sec:QQ24pt} introduces the chiral currents associated with VS and TS respectively. By modifying CLS free theory, we can obtain 4-point SYK-type terms on-shell. Following the same idea as we discussed in \cite{GHM}, 3-point and $q$-point superfield interactions will be introduced in section \ref{sec:3pt24pt} and \ref{sec:n-point} respectively. 4-point SYK-type vertices emerge in both cases when we go on-shell. Section \ref{sec:npt2npt} is devoted to the introduction of higher $q$-point superfield interactions which gives $2(q − 1)$-point fermionic interactions on-shell. Section \ref{sec:1D} shows the results for one dimensional Lagrangians that follow from the compactification of the Lagrangians constructed in four dimensions. The emergence of ${\cal N} = 4$ extended supersymmetry is made manifest. In section \ref{sec:dynamicalbosons} we explore the question, ``Do 1D, $\mathcal{N}=4$ SYK models necessarily require propagating bosons?" By studying two possibilities with CLS chiral current and Fayet-Iliopoulos mechanism, and the failure of these two models shows strong evidence that one cannot construct a 1D, ${\cal N}=4$ SYK model without dynamical bosons.
Finally, section \ref{sec:CONcL} gives conclusions. We follow the presentation of our work with an appendix and a bibliography.

\section{Review of 4D, $\cal N$ = 1 Theories in Two-Component Notation\label{sec:review}}

The derivations below use the same convention as the book {\it {Superspace}}. We list the convention as well as some useful identities in Appendix \ref{appen:convention}.
\subsection{Two-Component Notation CS}
Recall the superspace action for chiral supermultiplet (CS), which is
\begin{equation}
    \begin{split}
        {\cal L}_{\rm CS} ~=&~
         \int d^2\theta d^2\overline{\theta} ~ \Bar{\Phi}\Phi
        ~=~ \frac{1}{4}\,\rD^{\a}\rD_{\a}\Bar{\rD}^{\Dot \a}\Bar{\rD}_{\Dot\a}\,(\Bar{\Phi}\Phi) |    ~~~, \\
    \end{split}
\end{equation}
and the component fields are defined as
\begin{equation}
    A ~=~ \Phi| ~~,~~ \psi_{\a}~=~\rD_{\a}\Phi|~~,~~F~=~\rD^2\Phi|
~~~, \end{equation}
with propagating complex scalar bosonic $A$ and spinor fields $\psi_{\a}$.
One can then derive the D-equations that follow from these definitions.
\begin{align}
    \rD_{\a}\,A ~=~ \psi_{\a}~~,&~~\Bar{\rD}_{\dot\a}\,A ~=~ 0~~, \\
    \rD_{\a}\,\psi_{\b} ~=~ -C_{\a\b}\,F~~,&~~\Bar{\rD}_{\dot\a}\,\psi_{\b} ~=~ i\,\pa_{\b\dot\a}\,A~~, \\
    \rD_{\a}\,F ~=~ 0~~,&~~\Bar{\rD}_{\dot\a}\,F ~=~ i\,\pa^{\a}{}_{\dot\a}\,\psi_{\a}~~.
\end{align}

The Lagrangian in terms of component fields takes the form
\begin{equation}
    \begin{split}
        {\cal L}_{\rm CS} ~=&~ (\Box\Bar{A})A ~-~ i\,\psi_{\a}\,\pa^{\a\Dot\a}\Bar{\psi}_{\Dot\a} ~+~ F\Bar{F}
~~~.     \end{split}
\end{equation}

\subsection{Two-Component Notation CLS}

A complex linear supermultiplet (CLS) can be described by a complex linear superfield $\S$ which satisfy the constraint
\begin{equation}
    \Bar{\rD}^{2} \S ~=~ 0 ~~~.
\end{equation}
Its component fields are defined as
\begin{equation}
\begin{gathered}
    B ~=~ \S | ~~~, \\
    \rho_{\a}~=~\rD_{\a}\S| ~~~,~~~ \Bar{\zeta}_{\Dot\a}~=~\Bar{\rD}_{\Dot\a}\S| ~~~, \\
    H~=~\rD^2\S| ~~~,~~~ U_{\a\Dot\a}~=~\Bar{\rD}_{\Dot\a}\rD_{\a}\S| ~~~,~~~ \Bar{U}_{\a\Dot\a}~=~-\,{\rD}_{\a}\Bar{\rD}_{\Dot\a}\Bar{\S}|~~~, \\
    \Bar{\beta}_{\Dot\a} ~=~
    \frac{1}{2}\rD^{\a}\Bar{\rD}_{\Dot\a}\rD_{\a}\S| ~~~,
\end{gathered}
\end{equation}
with propagating complex scalar bosonic $B$ and and spinor fields
$\zeta_{\a}$.
One can then derive the D-equations that follow from these definitions.
\begin{align}
    &\rD_{\a}\,B ~=~ \rho_{\a}~~,~~\Bar{\rD}_{\dot\a}\,B ~=~ \Bar{\zeta}_{\dot\a}~~, \\
    &\rD_{\a}\,\rho_{\b} ~=~ -C_{\a\b}\,H~~,~~\Bar{\rD}_{\dot\a}\,\rho_{\b} ~=~ U_{\b\dot\a}~~, \\
    &\rD_{\a}\,\Bar{\zeta}_{\dot\b} ~=~ i\,\pa_{\a\dot\b}\,B -U_{\a\dot\b} ~~,~~\Bar{\rD}_{\dot\a}\,\Bar{\zeta}_{\dot\b} ~=~ 0~~, \\
    &\rD_{\a}\,H ~=~ 0~~,~~\Bar{\rD}_{\dot\a}\,H ~=~ \fracm{i}2\,\pa^{\a}{}_{\dot\a}\,\rho_{\a} - \Bar{\beta}_{\dot\a}~~,\\
    &\rD_{\a}\,U_{\b\dot\b} ~=~ i\pa_{\a\dot\b}\,\rho_{\b}+\fracm{i}{2}\,C_{\a\b}\pa^{\g}{}_{\dot\b}\rho_{\g}-C_{\a\b}\Bar{\beta}_{\dot\b}~~,~~\Bar{\rD}_{\dot\a}\,U_{\b\dot\b} ~=~ iC_{\dot\a\dot\b}\pa_{\b}{}^{\dot\g}\Bar{\zeta}_{\dot\g}~~,\\
   & \rD_{\a}\,\Bar{\b}_{\dot\b} ~=~ -\fracm{i}{2}\,\pa_{\a\dot\b}\,H ~~,~~\Bar{\rD}_{\dot\a}\,\Bar{\b}_{\dot\b} ~=~\fracm{i}{2}\pa_{\a\dot\b}\,U^{\a}{}_{\dot\a} +\pa^{\a}{}_{\dot\a}\pa_{\a\dot\b}B + i\,\pa^{\a}{}_{\dot\a}U_{\a\dot\b} ~~.
\end{align}

Recall the superspace action for complex linear supermuliplet, which is
\begin{equation}
    \begin{split}
        {\cal L}_{\rm CLS} ~=&~
         -\,\int d^2\theta d^2\overline{\theta} ~ \Bar{\S}\S
        ~=~ -\,\frac{1}{4}\,\rD^{\a}\rD_{\a}\Bar{\rD}^{\Dot \a}\Bar{\rD}_{\Dot\a}\,\{\,\Bar{\S}\S\,\}|\\
    \end{split}
\end{equation}
and this implies a component Lagrangian
\begin{equation}
    {\cal L}_{\rm CLS} ~=~ (\Box\Bar{B})B  ~-~ H\Bar{H} ~+~ \Bar{U}^{\a\Dot\a}U_{\a\Dot\a} ~-~ i\,\zeta_{\a}\,\pa^{\a\Dot\a}\Bar{\zeta}_{\Dot\a} ~+~ \beta^{\a}\rho_{\a} ~+~ \Bar{\beta}^{\Dot\a}\Bar{\rho}_{\Dot\a}
\end{equation}

One can modify the CLS constraint \cite{CNM} by introducing one more copy of the chiral superfield in the form of $\cq (\Phi)$,
\begin{equation}
    \Bar{\rD}^2\S~=~\cq(\Phi) ~~~,~~~ {\rD}^2\Bar{\S} ~=~\Bar{\cq}(\Bar\Phi) ~~~.
\label{eqn:QdefPhi}
\end{equation}
where $\cq (\Phi)$ is an arbitrary function of $\Phi$.
Direct calculations in the presence of this modification tell us
\begin{equation}
    \begin{split}
        {\rD}^2\Bar{\rD}^2 \Bar{\S} ~=&~ \Box\Bar{\S}~+~[\,\frac{1}{2}\,\Bar{\cq}''(\Bar{\rD}^{\Dot\a}\Bar{\Phi})(\Bar{\rD}_{\Dot\a}\Phi)~+~ \Bar{\cq}'(\Bar{\rD}^2 \Bar{\Phi})\,]~+~i\,\pa{}_{\a\Dot\a}\Bar{\rD}^{\Dot\a}\rD^{\a}\Bar{\S} ~~~, \\
        {\rD}^2\Bar{\rD}^2 {\S} ~=&~ \frac{1}{2}\,{\cq}''({\rD}^{\a}\Phi)({\rD}_{\a}{\Phi})~+~ {\cq}'({\rD}^2 {\Phi}) ~~~,
    \end{split}
\end{equation}
then the Lagrangian in terms of component fields is
\begin{equation}
     \begin{split}
        {\cal L}_{\rm CLS, \cq} ~=&~ (\Box\Bar{B})B  ~-~ H\Bar{H} ~+~ \Bar{U}^{\a\Dot\a}U_{\a\Dot\a} ~-~ i\,\zeta_{\a}\,\pa^{\a\Dot\a}\Bar{\zeta}_{\Dot\a} ~+~ \beta^{\a}\rho_{\a} ~+~ \Bar{\beta}^{\Dot\a}\Bar{\rho}_{\Dot\a} \\
        &~-~ \cq\Bar{\cq} ~+~ \Big\{~ -~ \cq' \Bar{B} F ~-~ \cq' \zeta^{\a} \psi_{\a} ~-~ \fracm12 \, \cq'' \Bar{B}\psi^{\a}\psi_{\a}  ~+~ {\rm h.\,c.} ~\Big\}
    \end{split}
\end{equation}

By introducing the $\cq(\Phi)$ field, we connect CLS with CS and we can move one step further to discuss their simplest dynamical relations.
Let's consider ${\cal L}_{\rm CS} + {\cal L}_{\rm CLS}$ and focus on the bosonic sector only,
\begin{equation}
    \begin{split}
        {\cal L}_B ~=&~ (\Box\Bar{A})A ~+~ F\Bar{F} ~+~ (\Box\Bar{B})B  ~-~ H\Bar{H} ~+~ \Bar{U}^{\a\Dot\a}U_{\a\Dot\a}\\
        &~-~ B\Bar{\cq}'\Bar{F}~-~\Bar{B}\cq' F ~-~ \Bar{\cq}\cq
    \end{split}
\end{equation}

Consider the on-shell Lagrangian by eliminating the auxiliary fields $F$, $H$, and $U_{\a\Dot\a}$.
\begin{equation}
    \begin{split}
        {\cal L}_B^{\rm on-shell} ~=&~ (\Box\Bar{A})A  ~+~ (\Box\Bar{B})B
       ~-~ \Bar{\cq}\cq
       ~-~ B\Bar{B}\cq'\Bar{\cq}'
    \end{split}
\end{equation}
If we take $\cq = -m\,\Phi$ with $m$ a positive parameter, the on-shell Lagrangian contains $-m^2(A\Bar{A}+B\Bar{B})$, indicating that these two fields have the same mass. If we turn to the fermionic sector of (2.3) and (2.19) using $\cq = - m\, \Phi$, then it is seen that the two-component fermions $\zeta_{\a}$ and ${\Bar \psi}_{\Dot \a}$ together form a massive Dirac fermion with the same mass as the
bosons.

\subsection{Two-Component Notation VS}

The quantity $V$ is an unconstrained real scalar superfield satisfying $V = \Bar{V}$ which defines component fields via
\begin{equation}
\begin{gathered}
    A_{\a \Dot\a} ~=~ \fracm12 \, [ \Bar{\rD}_{\Dot\a} , \rD_{\a} ] V| ~~~, \\
    \l_{\a} ~=~ i \Bar{\rD}^{2} \rD_{\a} V| ~~~,~~~
    \Bar{\l}_{\Dot\a} ~=~ - i \rD^{2} \Bar{\rD}_{\Dot\a} V| ~~~, \\
    d ~=~ \fracm12 \, \rD^{\a} \Bar{\rD}^{2} \rD_{\a} V| ~~~,
\end{gathered}
\end{equation}
which describes the vector supermultiplet (VS).
These are the only components which cannot be gauged away by non-derivative gauge transformations. In Wess-Zumino gauge, we retain these components only.

One can define the gauge invariant superfield, $ W_{\a} $, field strength via
\begin{equation}
    W_{\a} ~=~ i \Bar{\rD}^{2} \rD_{\a} V ~~~,~~~
    \Bar{W}_{\Dot\a} ~=~ - i \rD^{2} \Bar{\rD}_{\Dot\a} V ~~~,
\end{equation}
and one sees that it is chiral ($\Bar{\rD}_{\Dot\a} W_{\a}$ = 0). The components are
\begin{equation}
\begin{gathered}
    \l_{\a} ~=~ W_{\a} | ~~~, \\
    f_{\a\b} ~=~ \fracm12 \, \rD_{(\a} W_{\b)} | ~~~,~~~
    \Bar{f}_{\Dot\a\Dot\b} ~=~ -~ \fracm12 \, \Bar{\rD}_{(\Dot\a} \Bar{W}_{\Dot\b)} | ~~~,~~~
    d ~=~ -~ i \fracm12 \, \rD^{\a} W_{\a} |~=~  i \fracm12 \, \Bar{\rD}^{\Dot\a} \Bar{W}_{\Dot\a} |  ~~~, \\
    i \pa_{\a}{}^{\Dot\a} \Bar{\l}_{\Dot\a} ~=~ \rD^{2} W_{\a} | ~~~.
\end{gathered}
\end{equation}
One can then derive the D-equations that follow from these definitions.
\begin{align}
    &\rD_{\a}\,\l_{\b} ~=~ f_{\a\b}-iC_{\a\b}\,d~~,~~\Bar{\rD}_{\dot\a}\,\l_{\b} ~=~0~~, \\
    &\rD_{\a}\,f_{\b\g} ~=~ -\fracm{i}{2}\,C_{\a(\b}\pa_{\g)}{}^{\dot\a}\,\Bar{\l}_{\dot\a}~~,~~\Bar{\rD}_{\dot\a}\,f_{\b\g} ~=~ \fracm{i}{2}\,\pa_{(\b|\dot\a}\,\l_{|\g)}~~, \\
    &\rD_{\a}\,d ~=~ -\fracm12\,\pa_{\a}{}^{\dot\a}\,\Bar{\l}_{\dot\a}~~,~~\Bar{\rD}_{\dot\a}\,d ~=~ \fracm12\,\pa^{\b}{}_{\dot\a}\,\l_{\b} ~~.
\end{align}

Recall the superspace action for vector supermuliplet, which is
\begin{equation}
    \cl_{\rm VS} ~=~ \fracm14 \, \int d^{2} \theta ~ W^{\a} W_{\a} ~+~ {\rm h.\,c.}
\end{equation}
and the component Lagrangian is
\begin{equation}
    \cl_{\rm VS} ~=~ -~ \fracm14 \, f^{\a\b} f_{\a\b} ~-~ \fracm14 \, \Bar{f}^{\Dot\a\Dot\b} \Bar{f}_{\Dot\a\Dot\b}~-~ i \, \Bar{\l}^{\Dot\a} \pa_{\a\Dot\a} \l^{\a} ~+~ d^2
\end{equation}


To obtain
\ the familiar field strength $f_{\underaccent{\tilde}{a}\underaccent{\tilde}{b}}$, we have
\begin{equation}
    \begin{split}
        f_{\a\b} ~=&~ -\fracm12 \,\pa_{(\a}{}^{\Dot\a}\,A_{\b)\Dot\a} ~=~ -\fracm12\,\fracm{1}{\sqrt{2}}\,(\s^{\underaccent{\tilde}{a}\underaccent{\tilde}{b}})_{\a\b}\,f_{\underaccent{\tilde}{a}\underaccent{\tilde}{b}}\\
        f^{\dot\a\dot\b} ~=&~ \fracm12 \,\pa^{\a(\Dot\a}\,A_{\a}{}^{\Dot\b)} ~=~ \fracm12\,\fracm{1}{\sqrt{2}}\,(\s^{\underaccent{\tilde}{a}\underaccent{\tilde}{b}})^{\dot\a\dot\b}\,f_{\underaccent{\tilde}{a}\underaccent{\tilde}{b}}
    \end{split}
\end{equation}
where $f_{\a\b}$ and its conjugate is actually the self-dual and anti-self-dual parts of the field strength. Note that we have trace identities ($\epsilon^{0123}=1$)
\begin{equation}
    \begin{split}
        (\s^{\underaccent{\tilde}{a}\underaccent{\tilde}{b}})_{\a\b}(\s_{\underaccent{\tilde}{c}\underaccent{\tilde}{d}})^{\a\b} ~=&~ 2\,\d^{[\underaccent{\tilde}{a}}_{\underaccent{\tilde}{c}}\d^{\underaccent{\tilde}{b}]}_{\underaccent{\tilde}{d}} ~-~ 2i\,\epsilon^{\underaccent{\tilde}{a}\underaccent{\tilde}{b}}{}_{\underaccent{\tilde}{c}\underaccent{\tilde}{d}}\\
         (\s^{\underaccent{\tilde}{a}\underaccent{\tilde}{b}})_{\dot\a\dot\b}(\s_{\underaccent{\tilde}{c}\underaccent{\tilde}{d}})^{\dot\a\dot\b} ~=&~ 2\,\d^{[\underaccent{\tilde}{a}}_{\underaccent{\tilde}{c}}\d^{\underaccent{\tilde}{b}]}_{\underaccent{\tilde}{d}} ~+~ 2i\,\epsilon^{\underaccent{\tilde}{a}\underaccent{\tilde}{b}}{}_{\underaccent{\tilde}{c}\underaccent{\tilde}{d}}
    \end{split}
\end{equation}
Consequently we have
\begin{equation}
    \begin{split}
        f^{\a\b} f_{\a\b} ~+~ \Bar{f}^{\Dot\a\Dot\b} \Bar{f}_{\Dot\a\Dot\b} ~&=~ f_{\underaccent{\tilde}{a}\underaccent{\tilde}{b}} f^{\underaccent{\tilde}{a}\underaccent{\tilde}{b}} 
        \\
        f^{\a\b} f_{\a\b} ~-~ \Bar{f}^{\Dot\a\Dot\b} \Bar{f}_{\Dot\a\Dot\b} ~&=~ 
        -\,i\,\fracm 12 \, \epsilon^{\un a\un b\un c\un d}\,  f_{\un a\un b} f_{\un c\un d}  \\
    \end{split}
\end{equation}
and the bosonic sector of the Lagrangian reduces to the familiar Maxwell Lagrangian on- shell.

\subsection{Two-Component Notation TS}

The gauge tensor supermultiplet (TS) is described by a chiral spinor prepotential $\Phi_{\a}$ which satisfy $\Bar{\rD}_{\Dot\a} \Phi_{\a} = 0$ and has the following components,
\begin{equation}
\begin{gathered}
    \varphi ~=~ -~ \fracm12 \, \big(~ \rD^{\a} \Phi_{\a} ~+~ \Bar{\rD}^{\Dot\a} \Bar{\Phi}_{\Dot\a} ~\big) | ~~~,~~~
    b_{\a\b} ~=~ \fracm12 \, \rD_{(\a} \Phi_{\b)} | ~~~, \\
    \c_{\a} ~=~ \fracm12 \, \big(~ \rD^{2} \Phi_{\a} ~-~ i \pa_{\a}{}^{\Dot\a} \Bar{\Phi}_{\Dot\a} ~\big) | ~~~.
\end{gathered}
\end{equation}
Here we ignore the components that can be algebraically gauged away.

One can define a field strength superfield
\begin{equation}
    G ~=~ -~ \fracm12 \, \big(~ \rD^{\a} \Phi_{\a} ~+~ \Bar{\rD}^{\Dot\a} \Bar{\Phi}_{\Dot\a} ~\big)
\end{equation}
which is a real linear superfield, meaning $\rD^{2} G = \Bar{\rD}^{2} G = 0$. The components are
\begin{equation}
\begin{gathered}
    \varphi ~=~ G | ~~~, \\
    \c_{\a} ~=~ \rD_{\a} G | ~~~,~~~
    h_{\un{a}} ~=~ [ \Bar{\rD}_{\Dot\a} , \rD_{\a} ] G | ~~~.
\end{gathered}
\end{equation}
Here $h_{\un{a}}$ is a component field strength given by
\begin{equation}
    h_{\un{a}} ~=~ i \, \big(~ \pa^{\b}{}_{\Dot\a} b_{\a\b} ~-~ \pa_{\a}{}^{\Dot\b} \Bar{b}_{\Dot\a\Dot\b} ~\big) ~~~.
\end{equation}
One can then derive the D-equations that follow from these definitions.
\begin{align}
    &\rD_{\a}\,\varphi ~=~ \c_{\a}~~,~~\Bar{\rD}_{\dot\a}\,\varphi ~=~\Bar{\c}_{\dot\a}~~, \\
    &\rD_{\a}\,\c_{\b} ~=~ 0 ~~,~~\Bar{\rD}_{\dot\a}\,\c_{\b} ~=~ \fracm{i}{2}\,\pa_{\b\dot\a}\,\varphi + \fracm12\,h_{\b\dot\a}~~, \\
    &\rD_{\a}\,h_{\b\dot\b} ~=~ i\,\pa_{\a\dot\b}\,\c_{\b}- i\,C_{\a\b}\,\pa^{\g}{}_{\dot\b}\c_{\g}~~,~~\Bar{\rD}_{\dot\a}\,h_{\b\dot\b} ~=~ iC_{\dot\a\dot\b}\,\pa_{\b}{}^{\dot\g}\Bar{\c}_{\dot\g} - i\,\pa_{\b\dot\a}\,\Bar{\c}_{\dot\b} ~~.
\end{align}

The superspace Lagrangian is
\begin{equation}
    \cl_{\rm TS} ~=~ -~ \fracm12 \int d^{2} \theta d^{2}\overline{\theta} ~ G^{2} ~~~,
\end{equation}
and the corresponding component Lagrangian is
\begin{equation}
    \cl_{\rm TS} ~=~ \fracm14 \, \varphi \Box \varphi ~+~ \fracm14 \, h^{\un{a}} h_{\un{a}} ~-~ i \, \Bar{\c}^{\Dot\a} \pa_{\a\Dot\a} \c^{\a} ~~~.
\end{equation}

\newpage
\section{From Modified CLS Free Theory to 4-Point On-Shell SYK\label{sec:QQ24pt}}

We can construct chiral currents \cite{GHM}
\begin{align}
    \cj_{\rm VS} ~=&~ W^{\a} W_{\a} ~~~, \\
    \cj_{\rm TS} ~=&~ ( \Bar{\rD}^{\Dot\a} G ) ( \Bar{\rD}_{\Dot\a} G ) ~~~.
\end{align}
They all satisfy the chiral condition
\begin{equation}
    \Bar{\rD}_{\Dot\a} \cj ~=~ 0 ~~~.
\end{equation}
We can also define the cojugate (anti-chiral) currents 
\begin{align}
    \Bar{\cj}_{\rm VS} ~=&~ \Bar{W}^{\Dot\a} \Bar{W}_{\Dot\a} ~~~, \\
    \Bar{\cj}_{\rm TS} ~=&~ ( {\rD}^{\a} G ) ( {\rD}_{\a} G ) ~~~.
\end{align}
where $(\cj)^* = \Bar{\cj} $ and imply the equation $ {\rD}^{\Dot\a} \Bar{\cj}$ = 0. 

Recalling (\ref{eqn:QdefPhi})
\begin{equation}
    \Bar{\rD}^2\S~=~\cq(\Phi) ~~~,~~~ {\rD}^2\Bar{\S} ~=~\Bar{\cq}(\Bar\Phi)
    ~~~,
\end{equation}
One could modify the arguments of these $\cq$ functions 
\begin{equation}
    \cq_{\hca} (\Phi) ~\longrightarrow~ \cq_{\hca} (\cj) 
\end{equation}
with the form
\begin{equation}
    \cq_{\hca} (\cj) ~=~ \k_{\hca \cb \cc} \cj^{\cb \cc} 
\end{equation}
so that 
\begin{equation}
    \cq \Bar{\cq} ~\sim~ \k \k^{*} \cj \Bar{\cj} ~\sim~ \k \k^{*} \psi \psi \Bar{\psi} \Bar{\psi}
\end{equation}
and we obtain the 4-pt SYK-type terms!

Note that one can also define chiral currents from the CLS by constructing such currents from products of $(\Bar{\rD}^{\Dot\a}\Sigma^{\hcb}) $, i.e.
\begin{equation}
    \cj^{\hcb_{1}\hcb_{2}}_{\rm CLS} ~=~ (\Bar{\rD}^{\Dot\a}\Sigma^{\hcb_1}) (\Bar{\rD}_{\Dot\a}\Sigma^{\hcb_2}) ~~~,
\end{equation}
which satisfies the chiral condition
\begin{equation}
     \Bar{\rD}^{\Dot\b} \,\cj^{\hcb_{1}\hcb_{2}}_{\rm CLS} ~=~ 0 ~~~,
\end{equation}
when we choose $\cq = 0$ is the simplest such example. Therefore we will not use this current in the remainder of this chapter, as we will discuss the application of it in the next chapter.


\subsection{CLS,$\cq$(VS)}

\begin{equation}
    \begin{split}
        {\cal L}_{\rm CLS,\cq(VS)} ~=&~ 
         -\,\int d^2\theta d^2\overline{\theta} ~ \Bar{\S}^{\hca}\S_{\hca}
        ~=~ -\,\frac{1}{4}\,\rD^{\a}\rD_{\a}\Bar{\rD}^{\Dot \a}\Bar{\rD}_{\Dot\a}\,\{\,\Bar{\S}^{\hca}\S_{\hca}\,\}|\\
    \end{split}
\end{equation}
with 
\begin{equation}
    \Bar{\rD}^2\S_{\hca}~=~\k_{\hca \vcb \vcc} \cj_{\rm VS}^{\vcb \vcc}  ~~~,~~~ {\rD}^2\Bar{\S}_{\hca} ~=~\k^{*}_{\hca \vcb \vcc} \Bar{\cj}_{\rm VS}^{\vcb \vcc} 
\end{equation}

The Lagrangian in terms of component fields is
\begin{equation}
     \begin{split}
        {\cal L}_{\rm VS + CLS, \cq(VS)} ~=&~ -~ \fracm14 \, f^{\a\b \, \vca} f_{\a\b \, \vca} ~-~ \fracm14 \, \Bar{f}^{\Dot\a\Dot\b \,\vca} \Bar{f}_{\Dot\a\Dot\b\, \vca}~-~ i \, \Bar{\l}^{\Dot\a \, \vca} \pa_{\a\Dot\a} \l^{\a}_{\vca} ~+~ d^{\vca} d_{\vca}  \\
        &~+~ (\Box\Bar{B}^{\hca})B_{\hca}  ~-~ H^{\hca}\Bar{H}_{\hca} ~+~ \Bar{U}^{\hca\a\Dot\a}U_{\hca\a\Dot\a} ~-~ i\,\zeta^{\hca}_{\a}\,\pa^{\a\Dot\a}\Bar{\zeta}_{\hca\Dot\a} ~+~ \beta^{\hca\a}\rho_{\hca\a} ~+~ \Bar{\beta}^{\hca\Dot\a}\Bar{\rho}_{\hca\Dot\a} \\
        &~-~ \k^{\hce}{}_{\vca\vcb}\,\k^{*}_{\hce\vcc\vcd}\,\lambda^{\vca\a}\,\lambda_{\a}^{\vcb}\,\Bar{\lambda}^{\vcc\Dot\b}\,\Bar{\lambda}^{\vcd}_{\Dot\b} \\
        &~+~ \Big\{~ \k_{\hca\vcb\vcc} \, \Big[~
        i 2 \Bar{B}^{\hca} \l^{\a\vcb} \pa_{\a\Dot\a} \Bar{\l}^{\Dot\a \vcc} ~+~ \Bar{B}^{\hca} f^{\a\b \, \vcb} f_{\a\b}^{\vcc} ~-~ 2 \Bar{B}^{\hca} d^{\vcb} d^{\vcc} \\
        &~~~~~~~~ ~+~ 2 \z^{\a\hca} f_{\a\b}^{\vcb} \l^{\b\vcc} ~+~ i 2 \z^{\a\hca} d^{\vcb} \l_{\a}^{\vcc} ~\Big]
        ~+~ {\rm h.\,c.} ~\Big\} 
    \end{split}
\end{equation}

Although the off-shell Lagrangian already gives us the four-fermion SYK type term, the on-shell Lagrangian is still interesting, and it gives us more four-fermion coupling terms which involve fermions from the different multiplets.
\begin{equation}
    \begin{split}
         {\cal L}_{\rm VS + CLS, \cq(VS)}^{\rm on-shell} ~=&~ -\,\k^{\hce}{}_{\vca\vcb}\,\k^{*}_{\hce\vcc\vcd}\,\lambda^{\vca\a}\,\lambda_{\a}^{\vcb}\,\Bar{\lambda}^{\vcc\Dot\b}\,\Bar{\lambda}^{\vcd}_{\Dot\b}  \\
         &~+~ \Big\{\, \k_{\hca\vce\vcb}\,\k_{\hcc}{}^{\vce}{}_{\vcd}\,\zeta^{\hca\a}\,\lambda^{\vcb}_{\a}\,{\zeta}^{\hcc\b}\,{\lambda}^{\vcd}_{\b} ~-~ \k_{\hca\vce\vcb}\,\k^{*}_{\hcc}{}^{\vce}{}_{\vcd}\,\zeta^{\hca\a}\,\lambda^{\vcb}_{\a}\,\Bar{\zeta}^{\hcc\Dot\b}\,\Bar{\lambda}^{\vcd}_{\Dot\b}  ~+~ {\rm h.\,c.}\,\Big\} ~+~ \cdots
    \end{split}
\end{equation}

\subsection{CLS,$\cq$(TS)}

In this subsection, our starting point is
\begin{equation}
    \begin{split}
        {\cal L}_{\rm CLS,\cq(TS)} ~=&~ 
         -\,\int d^2\theta d^2\overline{\theta} ~ \Bar{\S}^{\hca}\S_{\hca}
        ~=~ -\,\frac{1}{4}\,\rD^{\a}\rD_{\a}\Bar{\rD}^{\Dot \a}\Bar{\rD}_{\Dot\a}\,\{\,\Bar{\S}^{\hca}\S_{\hca}\,\}|\\
    \end{split}
\end{equation}
with 
\begin{equation}
    \Bar{\rD}^2\S_{\hca}~=~\k_{\hca \tcb \tcc} \cj_{\rm TS}^{\tcb \tcc}  ~~~,~~~ {\rD}^2\Bar{\S}_{\hca} ~=~\k^{*}_{\hca \tcb \tcc} \Bar{\cj}_{\rm TS}^{\tcb \tcc} 
\end{equation}

The Lagrangian in terms of component fields is
\begin{equation}
     \begin{split}
        {\cal L}_{\rm TS + CLS, \cq(TS)} ~=&~ \fracm14 \, \varphi^{\tca} \Box \varphi_{\tca} ~+~ \fracm14 \, h^{\tca\un{a}} h_{\tca\un{a}} ~-~ i \, \Bar{\c}^{\tca\Dot\a} \pa_{\a\Dot\a} \c^{\a}_{\tca}  \\
        &~+~ (\Box\Bar{B}^{\hca})B_{\hca}  ~-~ H^{\hca}\Bar{H}_{\hca} ~+~ \Bar{U}^{\hca\a\Dot\a}U_{\hca\a\Dot\a} ~-~ i\,\zeta^{\hca}_{\a}\,\pa^{\a\Dot\a}\Bar{\zeta}_{\hca\Dot\a} ~+~ \beta^{\hca\a}\rho_{\hca\a} ~+~ \Bar{\beta}^{\hca\Dot\a}\Bar{\rho}_{\hca\Dot\a} \\
        &~-~ \k^{* \hce}{}_{\tca\tcb}\,\k_{\hce\tcc\tcd}\,\c^{\tca\a}\,\c_{\a}^{\tcb}\,\Bar{\c}^{\tcc\Dot\b}\,\Bar{\c}^{\tcd}_{\Dot\b} \\
        &~+~ \Big\{~ -\k_{\hca\tcb\tcc}\,\Big[\, -2\Bar{B}^{\hca}(i\pa_{\b\Dot\b}\chi^{\tcb\b})\,\Bar{\chi}^{\tcc\Dot\b} ~-~\fracm14\Bar{B}^{\hca}\Big(  i\pa_{\a\Dot\b}\varphi^{\tcb} - h_{\a\Dot\b}^{\tcb}\Big)\Big(  i\pa^{\a\Dot\b}\varphi^{\tcc} - h^{\tcc\a\Dot\b}\Big) \\
        &~-~\zeta^{\hca\a}\,\Big(  i\pa_{\a\Dot\b}\varphi^{\tcb} - h_{\a\Dot\b}^{\tcb}\Big)\,\Bar{\chi}^{\tcc\Dot\b} \,\Big]
        ~+~ {\rm h.\,c.} ~\Big\} 
    \end{split}
\end{equation}
The EoMs (i.e. 'equations of motion') for the auxiliary fields are trivial: all auxiliary fields vanish. So if we set $H^{\ca} = U_{\ca\a\Dot\a} = \beta^{\ca\a} = \rho^{\ca\a} = 0$ to obtain the on-shell Lagrangian.

\newpage
\section{From 3-Point Off-Shell Vertices to 4-Point On-Shell SYK}
\label{sec:3pt24pt}

\subsection{CS + CLS + 3PT-B}


Now, let us introduce a 3-point interaction between a complex anti-linear superfield and two chiral superfields. The full superfield Lagrangian is
\begin{equation}
\begin{split}
    \cl_{\rm CS+CLS+3PT-B} ~=~ \int d^2 \theta \, d^2 \Bar{\theta} ~ \Big[~
    \bBF^{\ca} \, \BF_{\ca} 
    ~-~  \bBS^{\hca} \, \BS_{\hca} ~\Big]
    ~+~ \Big\{~ \int d^2 \theta \, d^2 \Bar{\theta} ~ \Big[~ \Hat{\k}_{\hca\cb\cc} ~ \bBS^{\hca} \, \BF^{\cb} \, \BF^{\cc} ~\Big]
    ~+~ {\rm h.\, c.} ~\Big\}~~~,
\end{split}
\end{equation}
and the interaction term leads to the result
\begin{equation}
\begin{split}
    \cl_{\rm 3PT-B} ~=&~ \fracm{1}{4} \, {\Hat \k}_{\hca\cb\cc} ~ \rD^{\a}\rD_{\a}\Bar{\rD}^{\Dot \a}\Bar{\rD}_{\Dot\a}\, ~ \Big[~ 
    \bBS^{\hca} \, \BF^{\cb} \, \BF^{\cc} ~\Big] ~+~ {\rm h.\, c.}  \\
    ~=&~ \fracm{1}{2} \, {\Hat \k}_{\hca\cb\cc} ~ \rD^{\a}\rD_{\a} \,  \Big[ \, \big(  \Bar{\rD}^{2}\,\bBS^{\hca} \big) \, \BF^{\cb} \, \BF^{\cc} \,\Big] ~+~ {\rm {h.\, c.}} \\
    ~=&~ {\Hat \k}_{\hca\cb\cc}\,\Big\{\, -\,(\Box\Bar{B}^{\hca})A^\cb A^\cc~+~ \Bar{\cq}'^{\hca}{}_\cd(\Bar{A})\,A^\cb A^\cc \Bar{F}^\cd ~+~ i\,(\pa^{\a\Dot\a}\Bar{U}^{\hca}_{\a\Dot\a}) A^\cb A^\cc \\
    &~+~ \fracm12\,\Bar{\cq}''^{\hca}{}_{\cd\ce}(\Bar{A})\,A^\cb A^\cc \Bar{\psi}^{\cd \Dot\a}\Bar{\psi}^\ce_{\Dot\a} 
    ~+~i\,(\pa^{\a\Dot\a}\Bar{\rho}^{\hca}_{\Dot\a})\psi^{\cb}_{\a}A^\cc ~-~ 2\,\beta^{\hca\a}\psi^{\cb}_{\a}A^\cc \\
    &~+~ 2\,\Bar{H}^{\hca} F^{\cb} A^{\cc} ~+~ \Bar{H}^{\hca} \psi^{\cb \a}\psi^{\cc}_\a \,\Big\}~+~ {\rm {h.\, c.}} 
\end{split}
\end{equation}

The complete off-shell Lagrangian is
\begin{equation}
    \begin{split}
        \cl_{\rm CS+CLS+3PT-B}^{\rm off-shell} ~=&~ 
        (\Box\Bar{A}^{\ca})A_{\ca} ~+~ F^{\ca}\Bar{F}_{\ca} ~+~ (\Box\Bar{B}^{\hca})B_{\hca}  ~-~ H^{\hca}\Bar{H}_{\hca} ~+~ \Bar{U}^{\a\Dot\a}_{\hca}U^{\hca}_{\a\Dot\a}\\
        &~-~ B^{\hca}\Bar{\cq}'_{\hca\cb}(\Bar{A})\Bar{F}^{\cb}~-~\Bar{B}^{\hca}\cq'_{\hca\cb}(A) F^{\cb} ~-~ \Bar{\cq}^{\hca}(\Bar{A})\cq_{\hca}(A) \\
        &~-~ i\,\psi_{\a}^{\ca}\,\pa^{\a\Dot\a}\Bar{\psi}_{\ca\Dot\a}~-~ i\,\zeta^{\hca}_{\a}\,\pa^{\a\Dot\a}\Bar{\zeta}_{\hca\Dot\a} ~-~ \cq'\psi^{\a}\zeta_{\a} ~-~ \Bar{\cq}'^{\hca\cb}\Bar{\psi}_{\cb}^{\Dot\a}\Bar{\zeta}_{\hca\Dot\a} \\
        &~-~\frac{1}{2}\,\cq''^{\hca\cb\cc}(A)\psi^{\a}_{\cb}\psi_{\cc\a}\Bar{B}_{\hca}~-~\frac{1}{2}\,\Bar{\cq}''^{\hca\cb\cc}(\Bar{A})\,\Bar{\psi}_{\cb}^{\Dot\a}\Bar{\psi}_{\cc\Dot\a}{B}_{\hca} ~+~ \beta^{\hca\a}\rho_{\hca\a} ~+~ \Bar{\beta}^{\hca\Dot\a}\Bar{\rho}_{\hca\Dot\a} \\
        &~+~ \Big\{\,{\Hat \k}_{\hca\cb\cc}\,\Big[\, -\,(\Box\Bar{B}^{\hca})A^\cb A^\cc~+~ \Bar{\cq}'^{\hca}{}_\cd(\Bar{A})\,A^\cb A^\cc \Bar{F}^\cd ~-~ 2i\,\Bar{U}^{\hca}_{\a\Dot\a} (\pa^{\a\Dot\a}A^\cb) A^\cc \\
    &~+~  \fracm12\,\Bar{\cq}''^{\hca}{}_{\cd\ce}(\Bar{A})\,A^\cb A^\cc \Bar{\psi}^{\cd \Dot\a}\Bar{\psi}^\ce_{\Dot\a} 
    ~-~i\,\Bar{\rho}^{\hca}_{\Dot\a}(\pa^{\a\Dot\a}\psi^{\cb}_{\a})A^\cc~-~i\,\Bar{\rho}^{\hca}_{\Dot\a}\psi^{\cb}_{\a}(\pa^{\a\Dot\a}A^\cc) ~-~ 2\,\beta^{\hca\a}\psi^{\cb}_{\a}A^\cc \\
    &~+~ 2\,\Bar{H}^{\hca} F^{\cb} A^{\cc} ~+~ \Bar{H}^{\hca} \psi^{\cb \a}\psi^{\cc}_\a \,\Big]~+~ {\rm {h.\, c.}} \,\Big\}
    \end{split}
\end{equation}
When working through the EoMs for all of the auxiliary fields, one finds there is one interesting term whose leading contribution is a four-fermion SYK type term. 
\begin{equation}
    \begin{split}
        \cl_{\rm CS+CLS+3PT-B}^{\rm on-shell} ~=&~ \frac{{\Hat \k}^{\hca}{}_{\cb\cc}\,{\Hat \k}^{*\hcg}{}_{\cd\ce}}{\delta^{\hca\hcg}~+~\cy^{\hca\hcg}} \,\psi^{\cb\a}\,\psi^{\cc}_{\a}\,\Bar{\psi}^{\cd\Dot\a}\,\Bar{\psi}^{\ce}_{\Dot\a}
        ~+~ \cdots \\
        ~=&~ {\Hat \k}^{\hca}{}_{\cb\cc}\,{\Hat \k}^{*}_{\hca\cd\ce}\,\psi^{\cb\a}\,\psi^{\cc}_{\a}\,\Bar{\psi}^{\cd\Dot\a}\,\Bar{\psi}^{\ce}_{\Dot\a}  ~+~ \cdots \\
    \end{split}
\end{equation}
where 
\begin{equation}
    \cy^{\hca\hcg} ~=~ 4\,{\Hat \k}^{\hca\cb\cc}\,{\Hat \k}^{*\hcg}{}_{\cb\cd}\,A_{\cc}\,\Bar{A}^{\cd} ~~~.
\end{equation}

\newpage

\subsection{CS + CLS + 3PT-A + 3PT-B}
\label{subsec:N2susy}

Recalling the 3PT-A interaction 
\begin{equation}
\begin{split}
    \cl_{\rm 3PT-A} ~=&~ \int d^2 \theta \, d^2 \Bar{\theta} ~  \k_{\ca\cb\cc} \,\bBF^{\ca} \, \BF^{\cb} \, \BF^{\cc} ~+~ {\rm h.\, c.}\\
    ~=&~ \fracm{1}{4} \, \k_{\ca\cb\cc} \, \rD^{\a} \rD_{\a} \, \Bar{\rD}^{\Dot\b} \Bar{\rD}_{\Dot\b} ~ 
    \Big[~  \bBF^{\ca} \, \BF^{\cb} \, \BF^{\cc}  ~\Big]| ~+~ {\rm h.\, c.} \\
    ~=&~ {\k}_{\hca\cb\cc}\,\Big[\,(\Box\Bar{A}^{\hca})A^\cb A^\cc ~+~ 2\,(i\pa^{\a\Dot\a}\Bar{\psi}^{\ca}_{\Dot\a})\psi^{\cb}_{\a}A^{\cc} 
        ~+~ 2\,\Bar{F}^{\ca} F^{\cb} A^{\cc} ~+~ \Bar{F}^{\ca} \psi^{\cb \a}\psi^{\cc}_\a \,\Big]~+~ {\rm {h.\, c.}} ~~~,
\end{split}
\end{equation}
the complete off-shell Lagrangian is
\begin{equation}
    \begin{split}
        \cl_{\rm CS+CLS++3PT-A+3PT-B}^{\rm off-shell} ~=&~ 
        (\Box\Bar{A}^{\ca})A_{\ca} ~+~ F^{\ca}\Bar{F}_{\ca} ~+~ (\Box\Bar{B}^{\hca})B_{\hca}  ~-~ H^{\hca}\Bar{H}_{\hca} ~+~ \Bar{U}^{\a\Dot\a}_{\hca}U^{\hca}_{\a\Dot\a}\\
        &~-~ B^{\hca}\Bar{\cq}'_{\hca\cb}(\Bar{A})\Bar{F}^{\cb}~-~\Bar{B}^{\hca}\cq'_{\hca\cb}(A) F^{\cb} ~-~ \Bar{\cq}^{\hca}(\Bar{A})\cq_{\hca}(A) \\
        &~-~ i\,\psi_{\a}^{\ca}\,\pa^{\a\Dot\a}\Bar{\psi}_{\ca\Dot\a}~-~ i\,\zeta^{\hca}_{\a}\,\pa^{\a\Dot\a}\Bar{\zeta}_{\hca\Dot\a} ~-~ \cq'\psi^{\a}\zeta_{\a} ~-~ \Bar{\cq}'^{\hca\cb}\Bar{\psi}_{\cb}^{\Dot\a}\Bar{\zeta}_{\hca\Dot\a} \\
        &~-~\frac{1}{2}\,\cq''^{\hca\cb\cc}(A)\psi^{\a}_{\cb}\psi_{\cc\a}\Bar{B}_{\hca}~-~\frac{1}{2}\,\Bar{\cq}''^{\hca\cb\cc}(\Bar{A})\,\Bar{\psi}_{\cb}^{\Dot\a}\Bar{\psi}_{\cc\Dot\a}{B}_{\hca} ~+~ \beta^{\hca\a}\rho_{\hca\a} ~+~ \Bar{\beta}^{\hca\Dot\a}\Bar{\rho}_{\hca\Dot\a} \\
        &~+~\Big\{\,{\k}_{\hca\cb\cc}\,\Big[\,(\Box\Bar{A}^{\hca})A^\cb A^\cc ~+~ 2\,(i\pa^{\a\Dot\a}\Bar{\psi}^{\ca}_{\Dot\a})\psi^{\cb}_{\a}A^{\cc} \\
        &~+~ 2\,\Bar{F}^{\ca} F^{\cb} A^{\cc} ~+~ \Bar{F}^{\ca} \psi^{\cb \a}\psi^{\cc}_\a \,\Big]~+~ {\rm {h.\, c.}} \,\Big\}\\
        &~+~ \Big\{\,{\Hat \k}_{\hca\cb\cc}\,\Big[\, -\,(\Box\Bar{B}^{\hca})A^\cb A^\cc~+~ \Bar{\cq}'^{\hca}{}_\cd(\Bar{A})\,A^\cb A^\cc \Bar{F}^\cd ~-~ 2i\,\Bar{U}^{\hca}_{\a\Dot\a} (\pa^{\a\Dot\a}A^\cb) A^\cc \\
    &~+~ \fracm12\,\Bar{\cq}''^{\hca}{}_{\cd\ce}(\Bar{A})\,A^\cb A^\cc \Bar{\psi}^{\cd \Dot\a}\Bar{\psi}^\ce_{\Dot\a} 
    ~-~i\,\Bar{\rho}^{\hca}_{\Dot\a}(\pa^{\a\Dot\a}\psi^{\cb}_{\a})A^\cc~-~i\,\Bar{\rho}^{\hca}_{\Dot\a}\psi^{\cb}_{\a}(\pa^{\a\Dot\a}A^\cc) \\
    &~-~ 2\,\beta^{\hca\a}\psi^{\cb}_{\a}A^\cc 
    ~+~ 2\,\Bar{H}^{\hca} F^{\cb} A^{\cc} ~+~ \Bar{H}^{\hca} \psi^{\cb \a}\psi^{\cc}_\a \,\Big]~+~ {\rm {h.\, c.}} \,\Big\} ~~~,
    \end{split}
\end{equation}
after the complete derivation of the component results.

Deriving the equations of motion for auxiliary fields $F$ and $H$ yields 
\begin{align}
    F^{\ca} \, \Big[~ \d_{\cd\ca} ~+~ \cy_{\cd\ca} ~+~ \Bar{\cy}_{\ca\cd} ~\Big] ~+~ \Bar{\Hat{\cy}}_{\hcb\cd} \, H^{\hcb} ~=&~ -~ (\k\psi\psi)_{\cd} ~+~ \Bar{\cq}'_{\hcb\cd} \, \big( B ~-~ \Hat{\k} A A \big)^{\hcb} \\ 
    H_{\hca} ~=&~ \Hat{\cy}_{\hca\cb}F^{\cb} ~+~ (\Hat{\k}\psi\psi)_{\hca}
\end{align}
where the $\cy$'s are objects linear in $A$ or $\Bar{A}$ defined as
\begin{equation}
\begin{aligned}
    \cy_{\cd\ca} ~=&~ 2 \, \k_{\cd\ca\cc} \, A^{\cc} & ~,~~~
    \Bar{\cy}_{\cd\ca} ~=&~ 2 \, \k^{*}_{\cd\ca\cc} \, \Bar{A}^{\cc} ~~~, \\
    \Hat{\cy}_{\hcd\ca} ~=&~ 2 \, \Hat{\k}_{\hcd\ca\cc} \, A^{\cc} & ~,~~~
    \Bar{\Hat{\cy}}_{\hcd\ca} ~=&~ 2 \, \Hat{\k}^{*}_{\hcd\ca\cc} \, \Bar{A}^{\cc} ~~~,
\end{aligned}
\end{equation}
and the expressions with suppressed indices possess the definitions as
\begin{equation}
\begin{split}
    ( \k \psi \psi )_{\ca} ~=&~ \k_{\ca\cb\cc} \, \psi^{\a \cb} \psi_{\a}^{\cc} ~~~, \\
    ( \Hat{\k} \psi \psi )_{\hca} ~=&~ \Hat{\k}_{\hca\cb\cc} \, \psi^{\a \cb} \psi_{\a}^{\cc} ~~~, \\
    ( \Hat{\k} A A )_{\ca} ~=&~ \Hat{\k}_{\hca\cb\cc} \, A^{\cb} A^{\cc} ~~~.
\end{split}
\end{equation}
We note when expansion of the denominators, imply
\begin{equation}
\begin{split}
    \frac{1}{\d_{\ca\cb} ~+~ \cy_{\ca\cb}} ~=&~ \sum_{j=0}^{\infty} (-1)^j \big(\, \cy^{j} \,\big)^{\cb\ca} \\
    ~=&~ \d^{\cb\ca} ~-~ \cy^{\cb\ca} ~+~ \cy^{\cb\cc} \cy_{\cc}{}^{\ca} ~-~ \cy^{\cb\cc} \cy_{\cc\cd} \cy^{\cd\ca} ~+~ \cdots ~~~.
\end{split}
\end{equation}
and the altitude of the indices are lifted and the first and second indices flip. This is because we treat these as matrix multiplications and the R.H.S. consists of the inverses of both the identity matrix $\d_{\ca\cb}$ and the $\cy$-matrix $\cy_{\ca\cb}$, which are $\d^{\cb\ca}$ and $\cy^{\cb\ca}$ respectively.

Thus, we find
\begin{equation}
    \begin{split}
        \cl_{F,H}^{\rm on-shell} 
        ~=&~ (\Hat{\k}\psi\psi)^{\hca}(\Hat{\k}^*\Bar{\psi}\Bar{\psi})_{\hca} \\
        &~-~\frac{ \Big[\, (\k \psi\psi)_{\cb} + \Bar{\Hat{\cy}}_{\hce\cb} \, (\Hat{\k} \psi\psi)^{\hce} \,\Big] \, \left[ (\k^* \Bar{\psi}\Bar{\psi})_{\ca} + {\Hat{\cy}}_{\hcd\ca} \, (\Hat{\k}^* \Bar{\psi}\Bar{\psi})^{\hcd} \right]}{\d_{\cb\ca} ~+~ \cy_{\cb\ca} ~+~ \Bar{\cy}_{\ca\cb} ~+~ \Bar{\Hat{\cy}}^{\hcc}{}_{\cb} \Hat{\cy}_{\hcc\ca}}~+~ \mathcal{O}(\cq')  ~~~.
    \end{split}
\end{equation}
There are two terms whose leading contributions are four-fermion SYK type terms. The $\cl_{\rm F,H}$ sector is the only one that can contribute to SYK type terms. 
Explicitly, these leading terms are
\begin{equation}
    \begin{split}
        \cl_{\rm CS+CLS+3PT-A+3PT-B}^{\rm on-shell} 
        ~=&~ 
        -\,{\k}^{\ca}{}_{\cb\cc}\,{ \k}^{*}_{\ca\cd\ce}\,\psi^{\cb\a}\,\psi^{\cc}_{\a}\,\Bar{\psi}^{\cd\Dot\a}\,\Bar{\psi}^{\ce}_{\Dot\a}  ~+~ 
        {\Hat \k}^{\hca}{}_{\cb\cc}\,{\Hat \k}^{*}_{\hca\cd\ce}\,\psi^{\cb\a}\,\psi^{\cc}_{\a}\,\Bar{\psi}^{\cd\Dot\a}\,\Bar{\psi}^{\ce}_{\Dot\a}  ~+~ \cdots \\
    \end{split}
\end{equation}
Note that if we set $\k_{\ca\cb\cc} = \Hat{\k}_{\hca\cb\cc}$, all four-fermion terms vanish. We can see this by simply setting $\Omega_{\ca\cb}=\d_{\ca\cd}+\cy_{\ca\cb}$, then the on-shell $\cl_{\rm F,H}$ sector becomes
\begin{equation}
    \begin{split}
        \cl_{F,H}^{\rm on-shell} 
        ~=&~ ({\k}\psi\psi)^{\ca}({\k}^*\Bar{\psi}\Bar{\psi})_{\ca} 
        ~-~\frac{\Bar{\Omega}^{\cc\cb}\Omega^{\cd}{}_{\ca}}{\Bar{\Omega}^{\ce\cb}{\Omega}_{\ce\ca}}\,({\k}\psi\psi)_{\cc}({\k}^*\Bar{\psi}\Bar{\psi})_{\cd} ~+~ \mathcal{O}(\cq') \\
        ~=&~ ({\k}\psi\psi)^{\ca}({\k}^*\Bar{\psi}\Bar{\psi})_{\ca} 
        ~-~\Bar{\Omega}^{\cc\cb}\Omega^{\cd}{}_{\ca}(\Bar{\Omega}^{-1})_{\cb\ce}{(\Omega^{-1})}^{\ca\ce}\,({\k}\psi\psi)_{\cc}({\k}^*\Bar{\psi}\Bar{\psi})_{\cd}~+~ \mathcal{O}(\cq') \\
        ~=&~ ({\k}\psi\psi)^{\ca}({\k}^*\Bar{\psi}\Bar{\psi})_{\ca} 
        ~-~\d^{\cc}_{\ce}\d^{\cd\ce}\,({\k}\psi\psi)_{\cc}({\k}^*\Bar{\psi}\Bar{\psi})_{\cd}~+~ \mathcal{O}(\cq') ~=~ \mathcal{O}(\cq') \\
    \end{split}
\end{equation}

The vanishing of the solely four fermion terms for this choice $\k_{\ca\cb\cc} = \Hat{\k}_{\hca\cb\cc}$ is likely
a strong indication of an extended, but hidden, $\cn=2$ supersymmetry (in 4D, i.e. $\cn=8$ in 1D) of the model.  We make this assertion on
the basis of past results in the literature \cite{CNM,B&K,K1ps,CNM1,CNM2,CNM3,K4,CNM4,CNM4a,K5,K6,K7,K8}, wherein
it has been note that a pair of supermultiplets described by one chiral supermultiplet and one complex linear 
supermultiplet provides exactly the spectrum of component fields required to describe a Dirac spinor, two scalars
and two psuedoscalars of a 4D, $\cn=2$ supersymmetric hypermultiplet system.

\newpage
\section{From $q$-Point Off-Shell Vertices to 4-Point On-Shell SYK\label{sec:n-point}}

\subsection{CS + CLS +  nCLS-A}


An $n$-point superfield interaction among one chiral and a polynomial of complex linear superfields with kinetic terms can be written in the form
\begin{equation}
\begin{split}
    \cl_{\rm CS+CLS+nCLS-A} ~=~ \int d^2 \theta \, d^2 \Bar{\theta} ~ \Big\{~ 
    \bBF^{\ca} \, \BF_{\ca} 
    ~-~ \bBS^{\hca} \, \BS_{\hca} 
    ~+~\Big[~  \BF^{\ca} \cp_{\ca} (\BS)
    ~+~ {\rm h.\, c.}~\Big] \,\Big\} ~~~.
\end{split}
\end{equation}
The interaction term can be expressed as
\begin{equation}
    \cl_{\rm nCLS-A} ~=~ \fracm{1}{4} ~ \rD^{\a} \rD_{\a} \, \Bar{\rD}^{\Dot\a} \Bar{\rD}_{\Dot\a} ~ \Big[~ 
    \BF^{\ca} \cp_{\ca} (\BS) ~\Big] ~+~ {\rm h.\, c.}  ~~~.
\end{equation}
We let
\begin{equation}
    \cp_{\ca}(\BS) ~=~ \sum_{i=2}^{P} \k^{(i)}_{\ca\hcb_1\cdots\hcb_i} \prod_{k=1}^{i} \BS^{\hcb_k}  ~~~,
\label{eqn:PdefAAAA}
\end{equation}
where $\k^{(i)}_{\ca\hcb_1\cdots\hcb_{i}}$'s are arbitrary coefficients, and the degree of the polynomial is $P$. 
Note that the interaction terms start from cubic order.
Obviously, $\hcb_1$ to $\hcb_i$ indices for any $1\leq i \leq P$ on the coefficient $\k^{(i)}_{\ca\hcb_{1}\cdots\hcb_{i}}$ are symmetric.
We then have
\begin{align}
    \cp''_{\ca\hcb_1\hcb_2}(\BS) ~=&~ 2\, \k^{(2)}_{\ca\hcb_1\hcb_2}  ~+~ \sum_{j=3}^{P} j(j-1) \, \k^{(j)}_{\ca\hcb_1\cdots\hcb_{j}} \prod_{k=3}^{j} \BS^{\hcb_k}  ~~~,  \\
    \cp'''_{\ca\hcb_1\hcb_2\hcb_3}(\BS) ~=&~ 6\, \k^{(3)}_{\ca\hcb_1\hcb_2\hcb_3}  ~+~ \sum_{j=4}^{P} j (j-1) (j-2) \k^{(j)}_{\ca\hcb_1\cdots\hcb_{j}} \prod_{k=4}^{j} \BS^{\cb_k}  ~~~, \\
    \cp''''_{\ca\hcb_1\hcb_2\hcb_3\hcb_4}(\BS) ~=&~ 24\,\k^{(4)}_{\ca\hcb_1\hcb_2\hcb_3\hcb_4}  ~+~ \sum_{j=5}^{P} j (j-1) (j-2) (j-3) \k^{(j)}_{\ca\hcb_1\cdots\hcb_{j}} \prod_{k=5}^{j} \BS^{\hcb_k}  ~~~.
\end{align}
In terms of components, the complete off-shell Lagrangian is
\begin{equation}
    \begin{split}
        \cl_{\rm CS+CLS+nCLS-A}^{\rm off-shell} ~=&~ 
        (\Box\Bar{A}^{\ca})A_{\ca} ~+~ F^{\ca}\Bar{F}_{\ca} ~+~ (\Box\Bar{B}^{\hca})B_{\hca}  ~-~ H^{\hca}\Bar{H}_{\hca} ~+~ \Bar{U}^{\a\Dot\a}_{\hca}U^{\hca}_{\a\Dot\a}\\
        &~-~ B^{\hca}\Bar{\cq}'_{\hca\cb}(\Bar{A})\Bar{F}^{\cb}~-~\Bar{B}^{\hca}\cq'_{\hca\cb}(A) F^{\cb} ~-~ \Bar{\cq}^{\hca}(\Bar{A})\cq_{\hca}(A) \\
        &~-~ i\,\psi_{\a}^{\ca}\,\pa^{\a\Dot\a}\Bar{\psi}_{\ca\Dot\a}~-~ i\,\zeta^{\hca}_{\a}\,\pa^{\a\Dot\a}\Bar{\zeta}_{\hca\Dot\a} ~-~ \cq'\psi^{\a}\zeta_{\a} ~-~ \Bar{\cq}'^{\hca\cb}\Bar{\psi}_{\cb}^{\Dot\a}\Bar{\zeta}_{\hca\Dot\a} \\
        &~-~\frac{1}{2}\,\cq''^{\hca\cb\cc}(A)\psi^{\a}_{\cb}\psi_{\cc\a}\Bar{B}_{\hca}~-~\frac{1}{2}\,\Bar{\cq}''^{\hca\cb\cc}(\Bar{A})\,\Bar{\psi}_{\cb}^{\Dot\a}\Bar{\psi}_{\cc\Dot\a}{B}_{\hca} ~+~ \beta^{\hca\a}\rho_{\hca\a} ~+~ \Bar{\beta}^{\hca\Dot\a}\Bar{\rho}_{\hca\Dot\a} \\
        &~+~\Big\{\, F^{\ca}\,\Big[\,\fracm12\,\cp''_{\ca\hcb_1\hcb_2}(\BS)\Bar{\zeta}^{\hcb_1\Dot\a}\Bar{\zeta}^{\hcb_2}_{\Dot\a} ~+~ \cp'_{\ca\hcb_1}(\BS)\cq^{\hcb_1}(\BF) \,\Big]\\
        &~+~\psi^{\ca\a}\,\Big[\, \fracm12\,\cp'''_{\ca\hcb_1\hcb_2\hcb_3}(\BS)\rho^{\hcb_1}_{\a}\Bar{\zeta}^{\hcb_2\Dot\a}\Bar{\zeta}^{\hcb_3}_{\Dot\a} ~-~\cp''_{\ca\hcb_1\hcb_2}(\BS) \Bar{\zeta}^{\hcb_1\Dot\a}\left(i\pa_{\a\Dot\a}B^{\hcb_2}-U_{\a\Dot\a}^{\hcb_2}\right)\\
        &~~~~~~~~~~~~+~ \cp''_{\ca\hcb_1\hcb_2}(\BS)\rho_{\a}^{\hcb_1}\cq^{\hcb_2}(\BF) ~+~ \cp'_{\ca\hcb}(\BS)\cq'^{\hcb}{}_{\cc}\psi^{\cc}_{\a} \,\Big]\\
        &~+~ A^{\ca}\,\Big[\, \fracm14\,\cp''''_{\ca\hcb_1\hcb_2\hcb_3\hcb_4}(\BS) \rho^{\hcb_1\a}\rho^{\hcb_2}_{\a}\Bar{\zeta}^{\hcb_3\Dot\a}\Bar{\zeta}^{\hcb_4}_{\Dot\a} ~+~ \fracm12\,\cp'''_{\ca\hcb_1\hcb_2\hcb_3}(\BS)H^{\hcb_1}\Bar{\zeta}^{\hcb_2\Dot\a}\Bar{\zeta}^{\hcb_3}_{\Dot\a}\\
        &~~~~~~~~~~~~-~\cp'''_{\ca\hcb_1\hcb_2\hcb_3}(\BS) \rho^{\hcb_1}_{\a}\left(i\pa^{\a\Dot\a}B^{\hcb_2}-U^{\hcb_2\a\Dot\a}\right)\Bar{\zeta}^{\hcb_3}_{\Dot\a} \\
        &~~~~~~~~~~~~-~\fracm12\,\cp''_{\ca\hcb_1\hcb_2}(\BS)\left(i\pa^{\a\Dot\a}B^{\hcb_1}-U^{\hcb_1\a\Dot\a}\right)\left(i\pa_{\a\Dot\a}B^{\hcb_2}-U^{\hcb_2}_{\a\Dot\a}\right)\\
        &~~~~~~~~~~~~-~\cp''_{\ca\hcb_1\hcb_2}(\BS)\Bar{\zeta}^{\hcb_1\Dot\a}\Bar{\beta}^{\hcb_2}_{\Dot\a} ~+~ \fracm12\,\cp'''_{\ca\hcb_1\hcb_2\hcb_3}(\BS)\rho^{\hcb_1\a}\rho^{\hcb_2}_{\a}\cq^{\hcb_3}(\BS)\\
        &~~~~~~~~~~~~+~ \cp''_{\ca\hcb_1\hcb_2}(\BS)H^{\hcb_1}\cq^{\hcb_2}(\BS) ~-~ \cp''_{\ca\hcb_1\hcb_2}(\BS)\rho^{\hcb_1}_{\a}\cq'^{\hcb_2}{}_{\cc}(\BS)\psi^{\cc\a}\\
        &~~~~~~~~~~~~+~ \cp'_{\ca\hcb_1}(\BS)\left(\fracm12\,\cq''^{\hcb_1}{}_{\cc\cd}\psi^{\cc\a}\psi^{\cd}_{\a}+\cq'^{\hcb_1}{}_{\cc}F^{\cc} \right)
        \,\Big] \\
        &~-~ \frac{i}{2}\,\pa_{\a\Dot\a}\left(A^{\ca}\cp''_{\ca\hcb_1\hcb_2}(\BS)\Bar{\zeta}^{\hcb_1\Dot\a}\right)\rho^{\hcb_2\a}
        ~+~ {\rm {h.\, c.}} \,\Big\} ~~~.\\
    \end{split}
\end{equation}

For the on-shell lagrangian, there are two terms whose leading contributionss are four-fermion interaction terms. Explicitly these two leading terms are
\begin{equation}
    \begin{split}
         \cl_{\rm CS+CLS+nCLS-A}^{\rm on-shell} ~=&~ -\,\k^{(2) * \ca}{}_{\hcb_1\hcb_2}\,\k^{(2)}_{\ca\hcc_1\hcc_2}\,\zeta^{\hcb_1\a}\,\zeta^{\hcb_2}_{\a}\,\Bar{\zeta}^{\hcc_1\Dot\a}\,\Bar{\zeta}^{\hcc_2}_{\Dot\a} ~+~ 4\,\k^{(2)}_{\ca\hcb\hcc}\,\k^{(2)*}_{\cd\hce}{}^{\hcc}\,\psi^{\ca\a}\,\Bar{\zeta}^{\hcb\Dot\a}\,\Bar{\psi}^{\cd}_{\Dot\a}\,\zeta^{\hce}_{\a}~+~ \cdots
    \end{split}
\end{equation}

\newpage
\section{From $q$-Point Off-Shell Vertices to 2$(q-1)$-Point On-Shell SYK}
\label{sec:npt2npt}

\subsection{CS + CLS + nCLS-B}

Now consider the standard complex linear supermultiplet, we have 
\begin{equation}
    \Bar{\rD}^2\,\Sigma ~=~0 ~~~,
\end{equation}
therefore 
\begin{equation}
    \Bar{\rD}^{\Dot\b}\,\Big[\,(\Bar{\rD}^{\Dot\a}\Sigma) (\Bar{\rD}_{\Dot\a}\Sigma) \,\Big] ~=~ 2\,(\Bar{\rD}^{\Dot\b}\Bar{\rD}^{\Dot\a}\Sigma)(\Bar{\rD}_{\Dot\a}\Sigma) ~=~ -2\,(\Bar{\rD}^2\Sigma)(\Bar{\rD}^{\Dot\b}\Sigma) ~=~ 0 ~~~,
\end{equation}

Recall the current 
\begin{equation}
    \cj^{\hcb_{1}\hcb_{2}}_{\rm CLS} ~=~ (\Bar{\rD}^{\Dot\a}\Sigma^{\hcb_1}) (\Bar{\rD}_{\Dot\a}\Sigma^{\hcb_2}) ~~~,
\end{equation}
satisfies the chiral condition
\begin{equation}
     \Bar{\rD}^{\Dot\b} \,\cj^{\hcb_{1}\hcb_{2}}_{\rm CLS} ~=~ 0 ~~~.
\end{equation}
Also defining
\begin{equation}
    \Bar{\cj}^{\hcb_{1}\hcb_{2}}_{\rm CLS} ~=~  ({\rD}^{\a}\Bar{\Sigma}^{\hcb_1}) ({\rD}_{\a}\Bar{\Sigma}^{\hcb_2}) ~~~,
\end{equation}
and $(\cj)^{*} = \Bar{\cj}$.
The full superfield Lagrangian is
\begin{equation}
\begin{split}
    \cl_{\rm CS+CLS+nCLS-B} ~=&~ \int d^{2} \theta \, d^{2} \Bar{\theta} ~ \Big[~  \bBF^{\ca} \, \BF_{\ca} 
    ~-~ \bBS^{\hca} \BS_{\hca} ~\Big] \\
    &~+~ \Big\{~ \int d^{2} \theta ~ \Big[~  \BF^{\ca} \cf_{\ca} (\cj_{\rm CLS}) ~\Big] ~+~ {\rm h.\,c.} ~\Big\} ~~~,
\end{split}
\end{equation}
and the nCLS-B piece is
\be
    \cl_{\rm nCLS-B} ~=~ 
    \fracm12 \, {\rm D}^{\a}{\rm D}_{\a} ~ 
    \Big[~ \BF^{\ca} \cf_{\ca} (\cj_{\rm CLS}) ~\Big]  ~+~ {\rm {h.\, c.}}  ~~~,
\ee
where
\begin{align}
    \cf_{\ca} ( \cj_{\rm CLS} ) ~=&~ \sum_{i=1}^{P} \k^{(i)}_{\ca\hcb_{1}\hcb_{2}\cdots\hcb_{2i-1}\hcb_{2i}} \prod_{k=1}^{i} \cj^{\hcb_{2k-1}\hcb_{2k}}_{\rm CLS}  ~~~, \\
    \Bar{\cf}_{\ca} ( \Bar{\cj}_{\rm CLS} ) ~=&~ \sum_{i=1}^{P}  \, \k^{(i) *}_{\ca\hcb_{1}\hcb_{2}\cdots\hcb_{2i-1}\hcb_{2i}} \prod_{k=1}^{i} \Bar{\cj}^{\hcb_{2k-1}\hcb_{2k}}_{\rm CLS}  ~~~.
\end{align}
We obtain the off-shell component description as follows.
\begin{equation}
    \begin{split}
         \cl_{\rm CS+CLS+nCLS-B}^{\rm off-shell} ~=&~ 
        (\Box\Bar{A}^{\ca})A_{\ca} ~+~ F^{\ca}\Bar{F}_{\ca} ~+~ (\Box\Bar{B}^{\hca})B_{\hca}  ~-~ H^{\hca}\Bar{H}_{\hca} ~+~ \Bar{U}^{\a\Dot\a}_{\hca}U^{\hca}_{\a\Dot\a}\\
        &~-~ i\,\psi_{\a}^{\ca}\,\pa^{\a\Dot\a}\Bar{\psi}_{\ca\Dot\a}~-~ i\,\zeta^{\hca}_{\a}\,\pa^{\a\Dot\a}\Bar{\zeta}_{\hca\Dot\a}  ~+~ \beta^{\hca\a}\rho_{\hca\a} ~+~ \Bar{\beta}^{\hca\Dot\a}\Bar{\rho}_{\hca\Dot\a} \\
        &~+~ \Big\{\, F^{\ca}  \cf_{\ca} ( \cj_{\rm CLS} ) ~-~2\, \psi^{\ca\a}\, \cf'_{\ca\hcb_1\hcb_2} ( \cj_{\rm CLS} )\Bar{\zeta}^{\hcb_1\Dot\a}\,\left(i\pa_{\a\Dot\a}B^{\hcb_2}-U^{\hcb_2}_{\a\Dot\a}\right) \\
        &~+~ A^{\ca}\,\Big[\, 2\, \cf''_{\ca\hcb_1\hcb_2\hcb_3\hcb_4} ( \cj_{\rm CLS} )\Bar{\zeta}^{\hcb_1\Dot\a}\left(i\pa_{\a\Dot\a}B^{\hcb_2}-U^{\hcb_2}_{\a\Dot\a}\right)\Bar{\zeta}^{\hcb_3}_{\Dot\b}\left(i\pa^{\a\Dot\b}B^{\hcb_4}-U^{\hcb_4\a\Dot\b}\right)\\
        &~~~~~~~~~~~-~\fracm12\,\cf'_{\ca\hcb_1\hcb_2} ( \cj_{\rm CLS} )\left(i\pa^{\a\Dot\a}B^{\hcb_1}-U^{\hcb_1\a\Dot\a}\right)\left(i\pa_{\a\Dot\a}B^{\hcb_2}-U^{\hcb_2}_{\a\Dot\a}\right)\\
         &~~~~~~~~~~~-~ \cf'_{\ca\hcb_1\hcb_2} ( \cj_{\rm CLS} ) \Bar{\zeta}^{\hcb_1\Dot\a}\Bar{\beta}^{\hcb_2}_{\Dot\a}
        \,\Big] \\
        &~-~\frac{i}{2}\pa_{\a\Dot\a}\left( A^{\ca} \cf'_{\ca\hcb_1\hcb_2} ( \cj_{\rm CLS} )\Bar{\zeta}^{\hcb_1\Dot\a} \right)\,\rho^{\hcb_2\a}
         ~+~ {\rm {h.\, c.}} \,\Big\}
         ~~~.\\
    \end{split}
\end{equation}

The ${\cal L}_F$ sector of the on-shell Lagrangian clearly describes the 2$q$-fermion SYK type interactions 
\begin{equation}
    \begin{split}
       {\cal L}_F^{\rm on-shell} ~=&~   -\,\sum_{i,j=1}^{P} \k^{(i) \ca}{}_{\hcb_{1}\cdots\hcb_{2i}} \k^{(j)*}{}_{\ca\hcc_{1}\cdots\hcc_{2j}}\prod_{k=1}^{i} \prod_{l=1}^{j} \Bar{\z}^{\hcb_{2k-1}\Dot\a} \Bar{\z}_{\Dot\a}^{\hcb_{2k}} \z^{\hcc_{2l-1}\a} \z_{\a}^{\hcc_{2l}}
       ~~~.
    \end{split}
\end{equation}
Moreover, there is also a term comeing from the ${\cal L}_U$ sector of the on-shell lagrangian which is also purely fermionic.
\begin{equation}
    \begin{split}
        {\cal L}_U^{\rm on-shell} ~=&~ 
        4 \, \sum_{i,j=1}^{P} \k^{(i)}{}_{\ca\hcb}{}^{\hce}{}_{\hcf_3\cdots\hcf_{2i}} \k^{(j)*}{}_{\cc\hcd\hce\hcg_{3}\cdots\hcg_{2j}} \prod_{k=2}^{i} \prod_{l=2}^{j} \Bar{\z}^{\hcf_{2k-1}\Dot\b} \Bar{\z}_{\Dot\b}^{\hcf_{2k}} \z^{\hcg_{2l-1}\b} \z_{\b}^{\hcg_{2l}} ~ \psi^{\ca\a} \Bar{\z}^{\hcb\Dot\a} \Bar{\psi}^{\cd}_{\Dot\a} \z^{\hce}_{\a} ~~~. \\
    \end{split}
\end{equation}

\newpage
\section{1D, $\mathcal{N}=4$ SYK Models\label{sec:1D}} 
Given that in 4D,
\begin{equation}
    \begin{split}
         \pa_{\a\dot\a} ~=&~ (\s^{\underaccent{\tilde}{b}})_{\a\dot\a}\,\pa_{\underaccent{\tilde}{b}}~~,~~
    \pa_{\underaccent{\tilde}{b}} ~=~ \frac{1}{2}\,(\s_{\underaccent{\tilde}{b}})^{\a\dot\a}\,\pa_{\a\dot\a}
    \end{split}
\end{equation}
where the $\underaccent{\tilde}{b} = (\tau,x,y,z)$ labelling the 4D spacetime Cartesian coordinate 
and $(\s^{\underaccent{\tilde}{b}})^{\a\dot\a}$ is the standard Pauli matrix
representation with the zeroth component as the identity matrix, so numerically we have
\begin{equation}
    (\s^{\underaccent{\tilde}{0}})^{\a\dot\a} ~=~ \begin{pmatrix}
    1 & 0 \\
    0 & 1\\
    \end{pmatrix} ~:=~ \d^{\a\dot\a} ~~~,~~~
    (\s^{\underaccent{\tilde}{0}})_{\a\dot\a} ~=~ \begin{pmatrix}
    -1 & 0 \\
    0 & -1\\
    \end{pmatrix} ~:=~ -\,\d_{\a\dot\a} ~~~.
\end{equation}
where the metric is defined as
\begin{align}
    \eta_{\underaccent{\tilde}{a}\underaccent{\tilde}{b}}~=~\eta^{\underaccent{\tilde}{a}\underaccent{\tilde}{b}}~=~ {\rm diag}(-1,1,1,1)~~,~~
\end{align}
and the spinor metric is defined as
\begin{align}
    C_{\a\b} ~=~ C_{\dot\a\dot\b} ~=~ \begin{pmatrix}
    0 & -i\\
    i & 0
    \end{pmatrix} ~~,~~
    C^{\a\b} ~=~ C^{\dot\a\dot\b} ~=~ \begin{pmatrix}
    0 & i\\
    -i & 0
    \end{pmatrix} ~~.
\end{align}
One can obtain the following identities
\begin{equation}
    (\s_{\underaccent{\tilde}{b}})^{\a\dot\a}\,(\s^{\underaccent{\tilde}{c}})_{\a\dot\a} ~=~ 2\,\d_{\underaccent{\tilde}{b}}{}^{\underaccent{\tilde}{c}} ~~,~~ (\s^{\underaccent{\tilde}{b}})_{\a\dot\a}\,(\s_{\underaccent{\tilde}{b}})^{\b\dot\b} ~=~ 2\,\d_{\a}{}^{\b}\,\d_{\dot\a}{}^{\dot\b} ~~.
\end{equation}

When we carry out the dimension reduction, we have 
\begin{equation}
\pa_{\underaccent{\tilde}{a}} ~\rightarrow~
    \begin{cases}
    \pa_0 ~=~ \pa_{\tau}\\
    \pa_i ~=~ 0
    \end{cases} ~~,~~\pa^{\a\dot\a}~\rightarrow~
    \d^{\a\dot\a}\,\pa_{\tau}~~,~~
    \pa_{\a\dot\a}~\rightarrow~
    -\,\d_{\a\dot\a}\,\pa_{\tau}~~,~~
\end{equation}
and 
\begin{equation}
    \Box ~=~ \pa^{\underaccent{\tilde}{b}}\pa_{\underaccent{\tilde}{b}}~=~\frac{1}{2}\,\pa^{\a\dot\a}\,\pa_{\a\dot\a}~\rightarrow~-\,\pa_{\tau}\pa_{\tau} ~:=~ -\pa_{\t}^2 ~~.
\end{equation}
These rules suffice to obtain explicit results for component level expressions.  In our subsequent presentations, we utilize the equations above.


\subsection{CLS, $\cq$(VS)}

In the case of the CLS. $\cq$(VS) Lagrangian we find in 1D
\begin{equation}
     \begin{split}
        {\cal L}_{\rm VS + CLS, \cq(VS)} ~=&~ -~ \fracm14 \, f^{\a\b \, \vca} f_{\a\b \, \vca} ~-~ \fracm14 \, \Bar{f}^{\Dot\a\Dot\b \,\vca} \Bar{f}_{\Dot\a\Dot\b\, \vca}~+~ i \, \Bar{\l}^{\Dot\a \, \vca} \d_{\a\dot\a}\,\pa_{\tau} \l^{\a}_{\vca} ~+~ d^{\vca} d_{\vca}  \\
        &~+~ (\pa_{\tau}\Bar{B}^{\hca})(\pa_{\tau}B_{\hca})  ~-~ H^{\hca}\Bar{H}_{\hca} ~+~ \Bar{U}^{\hca\a\Dot\a}U_{\hca\a\Dot\a} ~-~ i\,\zeta^{\hca}_{\a}\,\d^{\a\dot\a}\,\pa_{\tau}\Bar{\zeta}_{\hca\Dot\a}   \\
        &~+~
       \beta^{\hca\a}\rho_{\hca\a} ~+~ \Bar{\beta}^{\hca\Dot\a}\Bar{\rho}_{\hca\Dot\a} ~-~ \k^{\hce}{}_{\vca\vcb}\,\k^{*}_{\hce\vcc\vcd}\,\lambda^{\vca\a}\,\lambda_{\a}^{\vcb}\,\Bar{\lambda}^{\vcc\Dot\b}\,\Bar{\lambda}^{\vcd}_{\Dot\b} \\
        &~+~ \Big\{~ \k_{\hca\vcb\vcc} \, \Big[~
       -\, i 2 \Bar{B}^{\hca} \l^{\a\vcb} \d_{\a\dot\a}\,\pa_{\tau} \Bar{\l}^{\Dot\a \vcc} ~+~ \Bar{B}^{\hca} f^{\a\b \, \vcb} f_{\a\b}^{\vcc} ~-~ 2 \Bar{B}^{\hca} d^{\vcb} d^{\vcc} \\
        &~~~~~~~~ ~+~ 2 \z^{\a\hca} f_{\a\b}^{\vcb} \l^{\b\vcc} ~+~ i 2 \z^{\a\hca} d^{\vcb} \l_{\a}^{\vcc} ~\Big]
        ~+~ {\rm h.\,c.} ~\Big\} ~~~.
    \end{split}
\end{equation}

\subsection{CLS, $\cq$(TS)}
In the case of the CLS. $\cq$(TS) Lagrangian we find in 1D
\begin{equation}
     \begin{split}
        {\cal L}_{\rm TS + CLS, \cq(TS)} ~=&~ \fracm14 \, (\pa_{\tau}\varphi^{\tca})\, (\pa_{\tau} \varphi_{\tca}) ~+~ \fracm14 \, h^{\tca\un{a}} h_{\tca\un{a}} ~+~ i \, \Bar{\c}^{\tca\Dot\a} \d_{\a\Dot\a}\pa_{\tau} \c^{\a}_{\tca}  \\
        &~+~ (\pa_{\tau}\Bar{B}^{\hca})(\pa_{\tau}B_{\hca})  ~-~ H^{\hca}\Bar{H}_{\hca} ~+~ \Bar{U}^{\hca\a\Dot\a}U_{\hca\a\Dot\a} ~-~ i\,\zeta^{\hca}_{\a}\,\d^{\a\Dot\a}\pa^{\tau}\Bar{\zeta}_{\hca\Dot\a} \\
        &~+~ \beta^{\hca\a}\rho_{\hca\a} ~+~ \Bar{\beta}^{\hca\Dot\a}\Bar{\rho}_{\hca\Dot\a} 
        ~-~ \k^{* \hce}{}_{\tca\tcb}\,\k_{\hce\tcc\tcd}\,\c^{\tca\a}\,\c_{\a}^{\tcb}\,\Bar{\c}^{\tcc\Dot\b}\,\Bar{\c}^{\tcd}_{\Dot\b} \\
        &~+~ \Big\{~ -\k_{\hca\tcb\tcc}\,\Big[\, 2i\,\Bar{B}^{\hca}\d_{\b\Dot\b}(\pa_{\tau}\chi^{\tcb\b})\,\Bar{\chi}^{\tcc\Dot\b}\\
        & ~~~~~~\,~~+~\fracm14\Bar{B}^{\hca}\Big(  i\d_{\a\Dot\b}\pa_{\tau}\varphi^{\tcb} + h_{\a\Dot\b}^{\tcb}\Big)\Big(  i\d^{\a\Dot\b}\pa_{\tau}\varphi^{\tcc} - h^{\tcc\a\Dot\b}\Big) \\
        &~+~\zeta^{\hca\a}\,\Big(  i\d_{\a\Dot\b}\pa_{\tau}\varphi^{\tcb} + h_{\a\Dot\b}^{\tcb}\Big)\,\Bar{\chi}^{\tcc\Dot\b} \,\Big]
        ~+~ {\rm h.\,c.} ~\Big\} ~~~.
    \end{split}
\end{equation}

\subsection{CS + CLS + 3PT-B} 
In the case of the CS + CLS + 3PT-B Lagrangian we find in 1D
\begin{equation}
    \begin{split}
        \cl_{\rm CS+CLS+3PT-B}^{\rm off-shell} ~=&~ 
        (\pa_{\tau}\Bar{A}^{\ca})(\pa_{\tau}A_{\ca}) ~+~ F^{\ca}\Bar{F}_{\ca} ~+~ (\pa_{\tau}\Bar{B}^{\hca})(\pa_{\tau}B_{\hca})  ~-~ H^{\hca}\Bar{H}_{\hca} ~+~ \Bar{U}^{\a\Dot\a}_{\hca}U^{\hca}_{\a\Dot\a}\\
        &~-~ B^{\hca}\Bar{\cq}'_{\hca\cb}(\Bar{A})\Bar{F}^{\cb}~-~\Bar{B}^{\hca}\cq'_{\hca\cb}(A) F^{\cb} ~-~ \Bar{\cq}^{\hca}(\Bar{A})\cq_{\hca}(A) \\
        &~-~ i\,\psi_{\a}^{\ca}\,\d^{\a\Dot\a}\pa^{\tau}\Bar{\psi}_{\ca\Dot\a}~-~ i\,\zeta^{\hca}_{\a}\,\d^{\a\Dot\a}\pa^{\tau}\Bar{\zeta}_{\hca\Dot\a} ~-~ \cq'\psi^{\a}\zeta_{\a} ~-~ \Bar{\cq}'^{\hca\cb}\Bar{\psi}_{\cb}^{\Dot\a}\Bar{\zeta}_{\hca\Dot\a} \\
        &~-~\frac{1}{2}\,\cq''^{\hca\cb\cc}(A)\psi^{\a}_{\cb}\psi_{\cc\a}\Bar{B}_{\hca}~-~\frac{1}{2}\,\Bar{\cq}''^{\hca\cb\cc}(\Bar{A})\,\Bar{\psi}_{\cb}^{\Dot\a}\Bar{\psi}_{\cc\Dot\a}{B}_{\hca} ~+~ \beta^{\hca\a}\rho_{\hca\a} ~+~ \Bar{\beta}^{\hca\Dot\a}\Bar{\rho}_{\hca\Dot\a} \\
        &~+~ \Big\{\,{\Hat \k}_{\hca\cb\cc}\,\Big[\, -2\,(\pa_{\tau}\Bar{B}^{\hca})(\pa_{\tau}A^\cb) A^\cc~+~ \Bar{\cq}'^{\hca}{}_\cd(\Bar{A})\,A^\cb A^\cc \Bar{F}^\cd ~-~ 2i\,\d^{\a\Dot\a}\Bar{U}^{\hca}_{\a\Dot\a} (\pa_{\tau}A^\cb) A^\cc \\
    &~+~  \fracm12\,\Bar{\cq}''^{\hca}{}_{\cd\ce}(\Bar{A})\,A^\cb A^\cc \Bar{\psi}^{\cd \Dot\a}\Bar{\psi}^\ce_{\Dot\a} 
    ~-~i\,\Bar{\rho}^{\hca}_{\Dot\a}\d^{\a\Dot\a}(\pa_{\tau}\psi^{\cb}_{\a})A^\cc~-~i\,\Bar{\rho}^{\hca}_{\Dot\a}\psi^{\cb}_{\a}\d^{\a\Dot\a}(\pa_{\tau}A^\cc)\\
    & ~-~ 2\,\beta^{\hca\a}\psi^{\cb}_{\a}A^\cc ~+~ 2\,\Bar{H}^{\hca} F^{\cb} A^{\cc} ~+~ \Bar{H}^{\hca} \psi^{\cb \a}\psi^{\cc}_\a \,\Big]~+~ {\rm {h.\, c.}} \,\Big\} ~~~.
    \end{split}
\end{equation}

\subsection{CS + CLS + 3PT-A + 3PT-B}
In the case of the
CS + CLS + 3PT-A + 3PT-B
Lagrangian, its 1D form is
\begin{equation}
    \begin{split}
        \cl_{\rm CS+CLS++3PT-A+3PT-B}^{\rm off-shell} ~=&~ 
        (\pa_{\t}\Bar{A}^{\ca})(\pa_{\t}A_{\ca}) ~+~ F^{\ca}\Bar{F}_{\ca} ~+~ (\pa_{\t}\Bar{B}^{\hca})(\pa_{\t}B_{\hca})  ~-~ H^{\hca}\Bar{H}_{\hca} ~+~ \Bar{U}^{\a\Dot\a}_{\hca}U^{\hca}_{\a\Dot\a}\\
        &~-~ B^{\hca}\Bar{\cq}'_{\hca\cb}(\Bar{A})\Bar{F}^{\cb}~-~\Bar{B}^{\hca}\cq'_{\hca\cb}(A) F^{\cb} ~-~ \Bar{\cq}^{\hca}(\Bar{A})\cq_{\hca}(A) \\
        &~-~ i\,\psi_{\a}^{\ca}\,\d^{\a\Dot\a}\,\pa_{\t}\Bar{\psi}_{\ca\Dot\a}~-~ i\,\zeta^{\hca}_{\a}\,\d^{\a\Dot\a}\,\pa_{\t}\Bar{\zeta}_{\hca\Dot\a} ~-~ \cq'\psi^{\a}\zeta_{\a} ~-~ \Bar{\cq}'^{\hca\cb}\Bar{\psi}_{\cb}^{\Dot\a}\Bar{\zeta}_{\hca\Dot\a} \\
        &~-~\frac{1}{2}\,\cq''^{\hca\cb\cc}(A)\psi^{\a}_{\cb}\psi_{\cc\a}\Bar{B}_{\hca}~-~\frac{1}{2}\,\Bar{\cq}''^{\hca\cb\cc}(\Bar{A})\,\Bar{\psi}_{\cb}^{\Dot\a}\Bar{\psi}_{\cc\Dot\a}{B}_{\hca} \\
        &~+~ \beta^{\hca\a}\rho_{\hca\a} ~+~ \Bar{\beta}^{\hca\Dot\a}\Bar{\rho}_{\hca\Dot\a} \\
        &~+~\Big\{\,{\k}_{\hca\cb\cc}\,\Big[\,2\,(\pa_{\t}\Bar{A}^{\hca})(\pa_{\t}A^\cb) A^\cc ~+~ 2\,(i\d^{\a\Dot\a}\,\pa_{\t}\Bar{\psi}^{\ca}_{\Dot\a})\psi^{\cb}_{\a}A^{\cc} \\
        &~+~ 2\,\Bar{F}^{\ca} F^{\cb} A^{\cc} ~+~ \Bar{F}^{\ca} \psi^{\cb \a}\psi^{\cc}_\a \,\Big]~+~ {\rm {h.\, c.}} \,\Big\}\\
        &~+~ \Big\{\,{\Hat \k}_{\hca\cb\cc}\,\Big[\,-2\, (\pa_{\t}\Bar{B}^{\hca})(\pa_{\t}A^\cb) A^\cc~+~ \Bar{\cq}'^{\hca}{}_\cd(\Bar{A})\,A^\cb A^\cc \Bar{F}^\cd \\
        &~-~ 2i\,\Bar{U}^{\hca}_{\a\Dot\a} (\d^{\a\Dot\a}\,\pa_{\t} A^\cb) A^\cc ~+~ \fracm12\,\Bar{\cq}''^{\hca}{}_{\cd\ce}(\Bar{A})\,A^\cb A^\cc \Bar{\psi}^{\cd \Dot\a}\Bar{\psi}^\ce_{\Dot\a} \\
    &~-~i\,\Bar{\rho}^{\hca}_{\Dot\a}(\d^{\a\Dot\a}\,\pa_{\t}\psi^{\cb}_{\a})A^\cc~-~i\,\Bar{\rho}^{\hca}_{\Dot\a}\psi^{\cb}_{\a}(\d^{\a\Dot\a}\,\pa_{\t}A^\cc) \\
    &~-~ 2\,\beta^{\hca\a}\psi^{\cb}_{\a}A^\cc 
    ~+~ 2\,\Bar{H}^{\hca} F^{\cb} A^{\cc} ~+~ \Bar{H}^{\hca} \psi^{\cb \a}\psi^{\cc}_\a \,\Big]~+~ {\rm {h.\, c.}} \,\Big\} ~~~.
    \end{split}
\end{equation}

\subsection{CS + CLS + nCLS-A}
In the case of the
CS + CLS + nCLS-A
Lagrangian, its 1D form is
\begin{equation}
    \begin{split}
        \cl_{\rm CS+CLS+nCLS-A}^{\rm off-shell} ~=&~ 
        (\pa_{\t}\Bar{A}^{\ca})(\pa_{\t}A_{\ca}) ~+~ F^{\ca}\Bar{F}_{\ca} ~+~ (\pa_{\t}\Bar{B}^{\hca})(\pa_{\t}B_{\hca})  ~-~ H^{\hca}\Bar{H}_{\hca} ~+~ \Bar{U}^{\a\Dot\a}_{\hca}U^{\hca}_{\a\Dot\a}\\
        &~-~ B^{\hca}\Bar{\cq}'_{\hca\cb}(\Bar{A})\Bar{F}^{\cb}~-~\Bar{B}^{\hca}\cq'_{\hca\cb}(A) F^{\cb} ~-~ \Bar{\cq}^{\hca}(\Bar{A})\cq_{\hca}(A) \\
        &~-~ i\,\psi_{\a}^{\ca}\,\d^{\a\Dot\a}\,\pa_{\t}\Bar{\psi}_{\ca\Dot\a}~-~ i\,\zeta^{\hca}_{\a}\,\d^{\a\Dot\a}\,\pa_{\t}\Bar{\zeta}_{\hca\Dot\a} ~-~ \cq'\psi^{\a}\zeta_{\a} ~-~ \Bar{\cq}'^{\hca\cb}\Bar{\psi}_{\cb}^{\Dot\a}\Bar{\zeta}_{\hca\Dot\a} \\
        &~-~\frac{1}{2}\,\cq''^{\hca\cb\cc}(A)\psi^{\a}_{\cb}\psi_{\cc\a}\Bar{B}_{\hca}~-~\frac{1}{2}\,\Bar{\cq}''^{\hca\cb\cc}(\Bar{A})\,\Bar{\psi}_{\cb}^{\Dot\a}\Bar{\psi}_{\cc\Dot\a}{B}_{\hca} ~+~ \beta^{\hca\a}\rho_{\hca\a} ~+~ \Bar{\beta}^{\hca\Dot\a}\Bar{\rho}_{\hca\Dot\a} \\
        &~+~\Big\{\, F^{\ca}\,\Big[\,\fracm12\,\cp''_{\ca\hcb_1\hcb_2}(\BS)\Bar{\zeta}^{\hcb_1\Dot\a}\Bar{\zeta}^{\hcb_2}_{\Dot\a} ~+~ \cp'_{\ca\hcb_1}(\BS)\cq^{\hcb_1}(\BF) \,\Big]\\
        &~+~\psi^{\ca\a}\,\Big[\, \fracm12\,\cp'''_{\ca\hcb_1\hcb_2\hcb_3}(\BS)\rho^{\hcb_1}_{\a}\Bar{\zeta}^{\hcb_2\Dot\a}\Bar{\zeta}^{\hcb_3}_{\Dot\a} ~+~\cp''_{\ca\hcb_1\hcb_2}(\BS) \Bar{\zeta}^{\hcb_1\Dot\a}\left(i\,\d_{\a\Dot\a}\,\pa_{\t} B^{\hcb_2}+U_{\a\Dot\a}^{\hcb_2}\right)\\
        &~~~~~~~~~~~~+~ \cp''_{\ca\hcb_1\hcb_2}(\BS)\rho_{\a}^{\hcb_1}\cq^{\hcb_2}(\BF) ~+~ \cp'_{\ca\hcb}(\BS)\cq'^{\hcb}{}_{\cc}\psi^{\cc}_{\a} \,\Big]\\
        &~+~ A^{\ca}\,\Big[\, \fracm14\,\cp''''_{\ca\hcb_1\hcb_2\hcb_3\hcb_4}(\BS) \rho^{\hcb_1\a}\rho^{\hcb_2}_{\a}\Bar{\zeta}^{\hcb_3\Dot\a}\Bar{\zeta}^{\hcb_4}_{\Dot\a} ~+~ \fracm12\,\cp'''_{\ca\hcb_1\hcb_2\hcb_3}(\BS)H^{\hcb_1}\Bar{\zeta}^{\hcb_2\Dot\a}\Bar{\zeta}^{\hcb_3}_{\Dot\a}\\
        &~~~~~~~~~~~~-~\cp'''_{\ca\hcb_1\hcb_2\hcb_3}(\BS) \rho^{\hcb_1}_{\a}\left(i\d^{\a\Dot\a}\,\pa_{\t}B^{\hcb_2}-U^{\hcb_2\a\Dot\a}\right)\Bar{\zeta}^{\hcb_3}_{\Dot\a} \\
        &~~~~~~~~~~~~+~\fracm12\,\cp''_{\ca\hcb_1\hcb_2}(\BS)\left(i\d^{\a\Dot\a}\,\pa_{\t}B^{\hcb_1}-U^{\hcb_1\a\Dot\a}\right)\left(i\d_{\a\Dot\a}\,\pa_{\t}B^{\hcb_2}+U^{\hcb_2}_{\a\Dot\a}\right)\\
        &~~~~~~~~~~~~-~\cp''_{\ca\hcb_1\hcb_2}(\BS)\Bar{\zeta}^{\hcb_1\Dot\a}\Bar{\beta}^{\hcb_2}_{\Dot\a} ~+~ \fracm12\,\cp'''_{\ca\hcb_1\hcb_2\hcb_3}(\BS)\rho^{\hcb_1\a}\rho^{\hcb_2}_{\a}\cq^{\hcb_3}(\BS)\\
        &~~~~~~~~~~~~+~ \cp''_{\ca\hcb_1\hcb_2}(\BS)H^{\hcb_1}\cq^{\hcb_2}(\BS) ~-~ \cp''_{\ca\hcb_1\hcb_2}(\BS)\rho^{\hcb_1}_{\a}\cq'^{\hcb_2}{}_{\cc}(\BS)\psi^{\cc\a}\\
        &~~~~~~~~~~~~+~ \cp'_{\ca\hcb_1}(\BS)\left(\fracm12\,\cq''^{\hcb_1}{}_{\cc\cd}\psi^{\cc\a}\psi^{\cd}_{\a}+\cq'^{\hcb_1}{}_{\cc}F^{\cc} \right)
        \,\Big] \\
        &~+~ \frac{i}{2}\,\d_{\a\Dot\a}\,\pa_{\t}\left(A^{\ca}\cp''_{\ca\hcb_1\hcb_2}(\BS)\Bar{\zeta}^{\hcb_1\Dot\a}\right)\rho^{\hcb_2\a}
        ~+~ {\rm {h.\, c.}} \,\Big\} ~~~. \\
    \end{split}
\end{equation}

\subsection{CS + CLS + nCLS-B}
In the case of the
CS + nCLS-B
Lagrangian, its 1D form is
\begin{equation}
    \begin{split}
        \cl_{\rm CS+CLS+nCLS-B}^{\rm off-shell} ~=&~ 
        (\pa_{\t} \Bar{A}^{\ca})(\pa_{\t} A_{\ca}) ~+~ F^{\ca}\Bar{F}_{\ca} ~+~ (\pa_{\t} \Bar{B}^{\hca})(\pa_{\t}B_{\hca})  ~-~ H^{\hca}\Bar{H}_{\hca} ~+~ \Bar{U}^{\a\Dot\a}_{\hca}U^{\hca}_{\a\Dot\a}\\
        &~-~ i\,\psi_{\a}^{\ca}\,\d^{\a\Dot\a}\,\pa_{\t}\Bar{\psi}_{\ca\Dot\a}~-~ i\,\zeta^{\hca}_{\a}\,\d^{\a\Dot\a}\,\pa_{\t}\Bar{\zeta}_{\hca\Dot\a}  ~+~ \beta^{\hca\a}\rho_{\hca\a} ~+~ \Bar{\beta}^{\hca\Dot\a}\Bar{\rho}_{\hca\Dot\a} \\
        &~+~ \Big\{\, F^{\ca}  \cf_{\ca} ( \cj_{\rm CLS} ) ~+~2\, \psi^{\ca\a}\, \cf'_{\ca\hcb_1\hcb_2} ( \cj_{\rm CLS} )\Bar{\zeta}^{\hcb_1\Dot\a}\,\left(i\,\d_{\a\Dot\a}\,\pa_{\t} B^{\hcb_2}+U^{\hcb_2}_{\a\Dot\a}\right) \\
        &~+~ A^{\ca}\,\Big[\, 2\, \cf''_{\ca\hcb_1\hcb_2\hcb_3\hcb_4} ( \cj_{\rm CLS} )\Bar{\zeta}^{\hcb_1\Dot\a}\left(\d_{\a\Dot\a}\,\pa_{\t}B^{\hcb_2}-iU^{\hcb_2}_{\a\Dot\a}\right)\Bar{\zeta}^{\hcb_3}_{\Dot\b}\left(\d^{\a\Dot\b} \pa_{\t} B^{\hcb_4}+U^{\hcb_4\a\Dot\b}\right)\\
        &~~~~~~~~~~~+~\fracm12\,\cf'_{\ca\hcb_1\hcb_2} ( \cj_{\rm CLS} )\left(i\d^{\a\Dot\a} \pa_{\t} B^{\hcb_1}-U^{\hcb_1\a\Dot\a}\right)\left(i\d_{\a\Dot\a}\pa_{\t} B^{\hcb_2}+U^{\hcb_2}_{\a\Dot\a}\right)\\
         &~~~~~~~~~~~-~ \cf'_{\ca\hcb_1\hcb_2} ( \cj_{\rm CLS} ) \Bar{\zeta}^{\hcb_1\Dot\a}\Bar{\beta}^{\hcb_2}_{\Dot\a}
        \,\Big] \\
        &~+~\frac{i}{2}\d_{\a\Dot\a}\pa_{\t}\left( A^{\ca} \cf'_{\ca\hcb_1\hcb_2} ( \cj_{\rm CLS} )\Bar{\zeta}^{\hcb_1\Dot\a} \right)\,\rho^{\hcb_2\a}
         ~+~ {\rm {h.\, c.}} \,\Big\}
         ~~~.\\
    \end{split}
\end{equation}
All the actionS of this section possess explicit 1D, $\cal N$ = 4 off-shell SUSY and SYK-type terms.

\newpage
\section{Evidence for the Incompatibility of SYK Terms and the Absence of Dynamical Bosons\label{sec:dynamicalbosons}}


One of the questions that occur when considering 1D, $\cal N$ = 4 extensions of SYK models
is, ``Do such model necessarily required propagating bosons?''  In the confines of the work
completed in \cite{GHM}, there were no concrete proofs related to this matter.  However,
the exploration of the diversity of models was undertaken with this question in mind.  The
results shown in that work covers all reasonable possibilities known to the authors in the attempt
to construct such models from the starting point of 4D, $\cal N$ = 1 supermultiplets.  The
results were negative in every studied case.

In unpublished work, we have also explored this question by starting with 2D, (4,0) heterotic
supermultiplets \cite{Adhm,hetN4b}. It is important to recognize that the smaller the number
of bosonic coordinates in a supermultiplet, the less restrictions result in the possible forms
of the supersymmetry transformation laws.  Nonetheless, even in these theories, all $\cal N$ = 4 
extensions of SYK models led to the presence of propagating bosons.  In the remainder of this
chapter, we will explore the same question.

\subsection{An attempt with CLS chiral current}

In this section, our starting point will include the 
possibility of using the complex linear supermultiplet as this has been the main topic of
this exploration.

Now, let us introduce an interaction between a complex anti-linear superfield and the CLS chiral current $\cj_{\rm CLS}$. This means that we have $\cq = 0$. The full superfield Lagrangian is
\begin{equation}
\begin{split}
    \cl_{\rm CLS+3PT-C} ~=&~ \int d^2 \theta \, d^2 \Bar{\theta} ~ \Big[
    ~-~ \Bar{\S}^{\hca} \, \S_{\hca}  ~\Big]
    ~+~ \Big\{~ \int d^2 \theta \, d^2 \Bar{\theta} ~ \Big[~ \k_{\hca\hcb\hcc} ~ \Bar{\S}^{\hca} \, \cj_{\rm CLS}^{\hcb\hcc} ~\Big] 
    ~+~ {\rm h.\, c.}  \,\Big\}  \\
    ~=&~ \int d^2 \theta \, d^2 \Bar{\theta} ~ \Big[
    ~-~ \Bar{\S}^{\hca} \, \S_{\hca}  ~\Big]
    ~+~ \Big\{~ \int d^2 \theta \, d^2 \Bar{\theta} ~ \Big[~ \k_{\hca\hcb\hcc} ~ \Bar{\S}^{\hca} \, (\Bar{\rD}^{\Dot\a} \S^{\hcb}) \, (\Bar{\rD}_{\Dot\a} \S^{\hcc}) ~\Big] 
    ~+~ {\rm h.\, c.} ~\Big\} ~~~.
\end{split}
\end{equation}
The complete off-shell Lagrangian is 
\begin{equation}
    \begin{split}
         \cl_{\rm CLS+3PT-C}^{\rm off-shell} ~=&~ 
          -\fracm12\,(\pa^{\g\dot\g}\Bar{B}^{\hca})(\pa_{\g\dot\g}B_{\hca})  ~-~ H^{\hca}\Bar{H}_{\hca} ~+~ \Bar{U}^{\a\Dot\a}_{\hca}U^{\hca}_{\a\Dot\a}
        ~-~ i\,\zeta^{\hca}_{\a}\,\pa^{\a\Dot\a}\Bar{\zeta}_{\hca\Dot\a}  ~+~ \beta^{\hca\a}\rho_{\hca\a} ~+~ \Bar{\beta}^{\hca\Dot\a}\Bar{\rho}_{\hca\Dot\a} \\
        & ~+~ \Big\{\, \k_{\hca\hcb\hcc}\,\Big[\, 
        ( \,\pa^{\g\dot\a}\Bar{B}^{\hca})\,(\pa_{\g\dot\g}\Bar{\zeta}^{\hcb\Dot\b})\Bar{\zeta}^{\hcc}_{\Dot\b} ~+~ i(\pa^{\a\Dot\a}\Bar{U}^{\hca}_{\a\Dot\a}) \Bar{\zeta}^{\hcb\Dot\b}\Bar{\zeta}^{\hcc}_{\Dot\b} \\
        &~~~~~~-~2\,\left(\fracm{i}{2}\pa^{\a\Dot\a}\Bar{\rho}^{\hca}_{\Dot\a} - \b^{\hca\a}\right)\Bar{\zeta}^{\hcb\Dot\b} \left(i\pa_{\a\Dot\b}B^{\hcc}-U_{\a\Dot\b}^{\hcc}\right)\\
        &~~~~~~-~\Bar{H}^{\hca}\left(i\pa^{\a\Dot\a}B^{\hcb}-U^{\hcb\a\Dot\a}\right)\left(i\pa_{\a\Dot\a}B^{\hcc}-U_{\a\Dot\a}^{\hcc}\right) \\
        &~~~~~+~ 2\,\Bar{H}^{\hca}\Bar{\zeta}^{\hcb\Dot\a}\left(\fracm{i}{2}\pa_{\a\Dot\a}\rho^{\hcc\a} - \Bar{\b}^{\hcc}_{\Dot\a}\right)
        \,\Big] ~+~ {\rm h.\, c.} ~\Big\} ~~~.
    \end{split}
\end{equation}
One sees that the all the terms involving the dynamical boson $B$ exist in the form of $\pa B$, and therefore one can do the replace $\pa B \rightarrow b$ ($b$ regarded as a new field) in 1D and think of it as an auxiliary field. That means we get rid of dynamical bosons in this model.

Now let us figure out if there exists a SYK term in this model. The Equations of Motions (EoMs) are:
\begin{align}
    \begin{split}
        H_{\hca} ~=&~ -\,\k_{\hca\hcb\hcc}\,\left(i\pa^{\a\Dot\a}B^{\hcb}-U^{\hcb\a\Dot\a}\right)\left(i\pa_{\a\Dot\a}B^{\hcc}-U_{\a\Dot\a}^{\hcc}\right) \\
        &~+~2\,\k_{\hca\hcb\hcc}\,\Bar{\zeta}^{\hcb\Dot\a}\left(\fracm{i}{2}\pa_{\a\Dot\a}\rho^{\hcc\a} - \Bar{\b}^{\hcc}_{\Dot\a}\right) ~~~,
    \end{split} \\
    \begin{split}
        U_{\hcc}^{\a\Dot\a} ~=&~ 2i\,\k_{\hcc\hcb\hcd}\,\Bar{\zeta}^{\hcb\Dot\b}(\pa_{\a\Dot\a}\Bar{\zeta}^{\hcd}_{\Dot\b})
        ~+~ 2i\,\k^*_{\hca\hcb\hcc}{H}^{\hca}(\pa^{\a\Dot\a}\Bar{B}^{\hcb}) \\
        & ~+~ 2\,\k^*_{\hca\hcb\hcc}\left(\fracm{i}{2}\pa^{\b\Dot\a}{\rho}_{\b}^{\hca}-\Bar{\beta}^{\hca\Dot\a}\right){\zeta}^{\hcb\a}~+~ 2\,\k^*_{\hca\hcb\hcc}{H}^{\hca}\Bar{U}^{\hcb\a\Dot\a}
        ~~~,
        \end{split} \\
    \begin{split}
        \Bar{\b}_{\hcc}^{\Dot\a} ~=&~ i\,\k_{\hcc\hcb\hcd}\,\pa^{\a\Dot\a}[\Bar{\zeta}^{\hcb\Dot\b}\left(i\pa_{\a\Dot\b}B^{\hcd}-U^{\hcd}_{\a\Dot\b}\right)] ~+~i\,\k^*_{\hca\hcb\hcc}\pa^{\a\Dot\a}[H^{\hca}\zeta^{\hcb}_{\a}] ~~~,
    \end{split} \\
    \rho_{\hcc\a} ~=&~ 2\,\k_{\hcc\hcb\hcd}\,\bar{\zeta}^{\hcb\Dot\a}\,\left(-i\,\pa_{\a\Dot\a}B^{\hcc} + U_{\a\Dot\a}^{\hcc}\right) ~+~ 2\,\k^*_{\hca\hcb\hcc}\,H^{\hca}\zeta_{\a}^{\hcb} ~~~.
\end{align}
A further calculation shows that 
\begin{align}
    \left(\fracm{i}{2}\pa_{\a\Dot\a}\rho^{\hcc\a} - \Bar{\b}^{\hcc}_{\Dot\a}\right) ~=&~ 2i\,\k^*_{\hca\hcb}{}^{\hcc}\,\pa_{\a\Dot\a}\left({H}^{\hca}{\zeta}^{\hcb\a}\right) ~~~,\\
    \left(\fracm{i}{2}\pa^{\b\Dot\a}{\rho}_{\b}^{\hca}-\Bar{\beta}^{\hca\Dot\a}\right) ~=&~ -2i\,\k^{\hca}{}_{\hcb\hcc}\,\pa^{\a\Dot\a}\left[\Bar{\zeta}^{\hcb\Dot\b}\left(i\pa_{\a\Dot\b}\Bar{B}^{\hcc}-\Bar{U}_{\a\Dot\b}^{\hcc}\right)\right] ~~~.
\end{align}
Therefore we have a system of equations in $H$ and $U$:
\begin{equation}
\begin{split}
    H_{\hcd} ~=&~ -~ i \, 4 \, \k_{\hcd\hcb\hcc} \, \k_{\hca}{}^{\hcb}{}_{\hce} \, \big[~ \pa^{\a\Dot\a} \big( H^{\hca} \z_{\a}^{\hce} \big) ~\big] \, \Bar{\z}_{\Dot\a}^{\hcc} \\
    &~-~ \k_{\hcd\hcb\hcc} \, \big( U_{\a\Dot\a}^{\hcb} ~-~ i \pa_{\a\Dot\a} B^{\hcb} \big) \, \big( U^{\a\Dot\a \, \hcc} ~-~ i \pa^{\a\Dot\a} B^{\hcc} \big) ~~~,
\end{split}
\end{equation}
\begin{equation}
\begin{split}
    \Bar{U}^{\a\Dot\a}_{\hcd} ~-~ 2 \, \k_{\hca\hcd\hcc} \Bar{H}^{\hca} U^{\a\Dot\a \, \hcc} ~=&~ -~ i \, 2 \, \k^{*}_{\hcd\hcb\hcc} \z^{\b \, \hcb} \, \big( \pa_{\a\Dot\a} \z_{\b}^{\hcc} \big) \\
    &~+~ i \, 4\, \k_{\hca\hcd\hcc} \, \k^{* \, \hca}{}_{\hcb\hce} \, \pa^{\a\Dot\b} \, \big[~ \big( \Bar{U}_{\b\Dot\b}^{\hcb} ~+~ i \pa_{\b\Dot\b} \Bar{B}^{\hcb} \big) \, \z^{\b \, \hce} ~\big] \, \Bar{\z}^{\Dot\a \, \hcc} \\
    &~-~ i \, 2 \, \k_{\hca\hcd\hcc} \, \Bar{H}^{\hca} \, \big( \pa^{\a\Dot\a} B^{\hcc} \big) ~~~,
\end{split}
\end{equation}
where we have $\pa H$ in EoM of $H$, and $\pa U$ in EoM of $U$, and they also couple. Thus the equations of motion become differential equations, and inverting these equations would definitely put derivatives on the fermion $\z$, which means it is not possible to obtain SYK terms from this model. 



\newpage
\subsection{An attempt to utilize the Fayet-Iliopoulos mechanism}

Another attempt would involve the Fayet-Iliopoulos mechanism.
The Fayet-Iliopoulos term here is a linear term in the ``spectator'' chiral superfield. When one does the chiral superspace integral, one would obtain a linear auxiliary field term. Therefore, if one sticks the auxiliary field to any term we want (e.g. a quartic fermionic SYK term here) in the off-shell Lagrangian, one would obtain the desired term in the on-shell Lagrangian as the equation of motion of that auxiliary field would contain a constant and the SYK term. For example, one could have
\begin{equation}
\begin{split}
    \cl_{\rm FI-1} ~=&~ \int d^2 \theta d^2 \Bar{\theta} ~ \Big[~ \Phi \Bar{\Phi} ~+~ \Big(~ \fracm14 \, W^{\a \vca} W_{\a \vca} ~+~ {\rm h.\,c.} ~\Big) ~\Big] \\
    &~+~ \Big\{~ \int d^2 \theta ~ c \, \Phi ~+~ {\rm h.\,c.} ~\Big\} \\
    &~+~ \Big\{~ \int d^2 \theta ~ \Big[~ \tilde{\k}_{\vca\vcb\vcc\vcd} \, \Phi \, W^{\a \vca} W_{\a}^{\vcb} W^{\b \vcc} W_{\b}^{\vcd} ~\Big] ~+~ {\rm h.\,c.} ~\Big\} ~~~,
\end{split}
\end{equation}
since the VS field strength $W_{\a}$ is chiral. Note that the ``spectator'' chiral superfield $\Phi$ does not have a number of copies index. So this action would lead to an off-shell Lagrangian with
\begin{equation}
    \cl^{\rm off-shell}_{\rm FI-1} ~=~ F \Bar{F} ~+~ \Big\{~ c F ~+~ {\rm h.\,c.} ~\Big\} ~+~ \Big\{~ \tilde{\k}_{\vca\vcb\vcc\vcd} \, F \, \l^{\a \vca} \l_{\a}^{\vcb} \l^{\b \vcc} \l_{\b}^{\vcd} ~+~ {\rm h.\,c.} ~\Big\} ~+~ \cdots ~~~,
\end{equation}
and the equation of motion for $F$ would be
\begin{equation}
    \Bar{F} ~+~ c ~+~ \l^{\a \vca} \l_{\a}^{\vcb} \l^{\b \vcc} \l_{\b}^{\vcd} ~=~ 0 ~~~.
\end{equation}
Therefore, the kinetic and FI terms would generate our desired on-shell term
\begin{equation}
    \cl^{\rm on-shell}_{\rm FI-1} ~\sim~ \Bar{c} \, \k \, \l \l \l \l ~+~ {\rm h.\,c.} ~+~ \cdots ~~~.
\end{equation}
However, the problem of this action is that it would generate terms with dynamical bosons,
\begin{equation}
    \cl_{\rm FI-1} ~=~ \tilde{\k}_{\vca\vcb\vcc\vcd} \, \Phi \, \rD^{2} \Big[~ W^{\a \vca} W_{\a}^{\vcb} W^{\b \vcc} W_{\b}^{\vcd} ~\Big] ~+~ {\rm h.\,c.} ~+~ \cdots ~~~,
\end{equation}
which is what we want to get rid of. Note that this action actually resembles the nVS-B model in \cite{GHM}.

Now, we note that $W_{\a}$ is chiral, and therefore modifying this interaction term by changing $\Phi$ to $\Bar{\Phi}$ (which allows us to integrate the entire superspace) would put derivatives only on $\Bar{\Phi}$. This would allow us to change all $\Phi$'s to $\Bar{\rD}^2 \Bar{\Phi}$ and get rid of all dynamical bosons. 
\begin{equation}
\begin{split}
    \cl_{\rm FI-2} ~=&~ \int d^2 \theta d^2 \Bar{\theta} ~ \Big[~ \Phi \Bar{\Phi} ~+~ \Big(~ \fracm14 \, W^{\a \vca} W_{\a \vca} ~+~ {\rm h.\,c.} ~\Big) ~\Big] \\
    &~+~ \Big\{~ \int d^2 \theta ~ c \, \Phi ~+~ {\rm h.\,c.} ~\Big\} \\
    &~+~ \Big\{~ \int d^2 \theta d^2 \Bar{\theta} ~ \Big[~ \tilde{\k}_{\vca\vcb\vcc\vcd} \, \Bar{\Phi} \, W^{\a \vca} W_{\a}^{\vcb} W^{\b \vcc} W_{\b}^{\vcd} ~\Big] ~+~ {\rm h.\,c.} ~\Big\} ~~~.
\end{split}
\end{equation}
However, the linear term cannot be changed accordingly as it is chiral and cannot be integrated over the entire superspace. Therefore we lose the 'magic power' of the Fayet-Iliopoulos mechanism - that is to utilize the constant $c$ in the equation of motion of the auxiliary field to obtain any term we desire. Nonetheless, we can still try to find its off-shell component Lagrangian and take a look at the equations of motion to see if we have any luck.
The complete off-shell Langrangian is
\begin{equation}
    \begin{split}
        {\cal L}_{\rm FI-2}^{\rm off-shell} ~=&~ -\fracm14\, f^{\vca}_{\underaccent{\tilde}{a}\underaccent{\tilde}{b}} f_{\vca}^{\underaccent{\tilde}{a}\underaccent{\tilde}{b}} - i\,\Bar{\lambda}_{\vca}^{\dot\a}\,\pa_{\a\dot\a}\,\lambda^{\vca\a} + d^{\vca}\,d_{\vca} -\fracm12\,(\pa_{\a\dot\a}\Bar{A})(\pa^{\a\dot\a}A) - i\,\psi_{\a}\,\pa^{\a\dot\a}\,\Bar{\psi}_{\dot\a} + F\Bar{F}\\
        &+ cF + \bar{c}\Bar{F} + \Big\{\, 
        4\,\tilde{\k}_{\vca\vcb\vcc\vcd} \,\Bar{F}\,\Big[\, i\,(\pa^{\a\dot\a}\,\Bar{\lambda}^{\vca}_{\dot\a})\,\lambda^{\vcb}_{\a}\,\lambda^{\vcc\b}\,\lambda^{\vcd}_{\b} - \fracm12\,f^{\vca}_{\a\b}\,f^{\vcb\a\b}\,\lambda^{\vcc\g}\,\lambda^{\vcd}_{\g} \\
        &\qquad\qquad\qquad\qquad\qquad\qquad + d^{\vca}\,d^{\vcb}\,\lambda^{\vcc\a}\,\lambda^{\vcd}_{\a} + f^{\vca}_{\a\b}\,\lambda^{\vcb\b}\,f^{\vcc\a\g}\,\lambda^{\vcd}_{\g}\\
        &\qquad\qquad\qquad\qquad\qquad\qquad -2i\,f^{\vca}_{\a\b}\,\lambda^{\vcb\b}\,d^{\vcc}\,\lambda^{\vcd\a} - d^{\vca}\,\lambda^{\vcb\a}\,d^{\vcc}\,\lambda^{\vcd\a} \,\Big] \\
        &\qquad\qquad\qquad - 2\,\tilde{\k}_{\vca\vcb\vcc\vcd} \,(\pa_{\g\dot\g}\,\Bar{A})\,(\pa^{\g\dot\g}\,\lambda^{\vca\a})\,\lambda^{\vcb}_{\a}\,\lambda^{\vcc\b}\,\lambda^{\vcd}_{\b} \\
        &\qquad\qquad\qquad - 4i\,\tilde{\k}_{\vca\vcb\vcc\vcd}\,(\pa^{\g\dot\g}\,\Bar{\psi}_{\dot\g})\,\Big[\,f^{\vca}_{\g\a}\,\lambda^{\vcb\a}\, + i\,d^{\vca}\,\lambda^{\vcb}_{\g}\,\Big] \,\lambda^{\vcc\b}\,\lambda^{\vcd}_{\b}
        + {\rm h.~c.} \,\Big\} ~~~.
    \end{split}
\end{equation}
Notice that after using the EoMs for the auxiliary field $F$
\begin{equation}
    \begin{split}
        F ~=&~ -\,\bar{c} - 4\,\tilde{\k}_{\vca\vcb\vcc\vcd} \,\Big[\, i\,(\pa^{\a\dot\a}\,\Bar{\lambda}^{\vca}_{\dot\a})\,\lambda^{\vcb}_{\a}\,\lambda^{\vcc\b}\,\lambda^{\vcd}_{\b} 
       + d^{\vca}\,d^{\vcb}\,\lambda^{\vcc\a}\,\lambda^{\vcd}_{\a}  - d^{\vca}\,\lambda^{\vcb\a}\,d^{\vcc}\,\lambda^{\vcd\a} \,\Big]  + \mathcal{O}(f_{\a\b}) ~~~, \\
    \end{split}
\end{equation}
and its conjugate, there are only three terms contributed by $d$, $F$, and $\Bar{F}$ fields in the Lagrangian
\begin{equation}
    \begin{split}
        \cl_{\rm FI-2} ~\sim~ d^{\vca}\,d_{\vca}
- F\Bar{F} + \Big[\, 4\,\tilde{\k}_{\vca\vcb\vcc\vcd}\,(\pa^{\g\dot\g}\,\Bar{\psi}_{\dot\g})\,d^{\vca}\,\lambda^{\vcb}_{\g}\,\lambda^{\vcc\b}\,\lambda^{\vcd}_{\b}+ {\rm h.~c.} \,\Big] ~~~.
    \end{split}
\end{equation}
Only when the leading term in the solution of the EoM for $d$ field $\sim$ $\lambda$ or $\lambda\,\lambda$, we can obtain SYK-type terms in the final on-shell Lagrangian. However, the EoM for $d$ takes the following form
\begin{equation}
    \begin{split}
        d_{\vca} ~=&~ -4\,\tilde{\k}_{\vca\vcb\vcc\vcd}\,\Bar{F}\,d^{\vcb}\,\lambda^{\vcc\a}\,\lambda^{\vcd}_{\a} 
         +4\,\tilde{\k}_{\vca\vcb\vcc\vcd}\,\Bar{F}\,\lambda^{\vcb\a}\,d^{\vcc}\,\lambda^{\vcd}_{\a} - 2\,\tilde{\k}_{\vca\vcb\vcc\vcd}\,(\pa^{\g\dot\g}\,\Bar{\psi}_{\dot\g})\,\lambda^{\vcb}_{\g}\,\lambda^{\vcc\b}\,\lambda^{\vcd}_{\b} + \mathcal{O}(f_{\a\b}) \\
        &~+~ {\rm h.~c.} ~~~,
    \end{split}
\end{equation}
where we can substitute (8.18) and its conjugate into the above equation and get a cubic order equation for $d$. 
So we obtain a conclusion that only the equation of motion of $F$ contains pure $\l$-terms, however, it takes the form of 
\begin{equation}
    F ~=~ -\,\bar{c} - i 4\,\tilde{\k}_{\vca\vcb\vcc\vcd} \, (\pa^{\a\dot\a}\,\Bar{\lambda}^{\vca}_{\dot\a})\,\lambda^{\vcb}_{\a}\,\lambda^{\vcc\b}\,\lambda^{\vcd}_{\b} ~+~ \cdots ~~~,
\end{equation}
which inevitably has a derivative on $\l$. Therefore, neither the kinetic term nor the FI-term would generate purely fermionic terms with no derivatives acting on any of the fermions. 

We could also consider the adinkra duality trick of $\pa A \rightarrow b$ in 1D. The equation of motion of the new auxiliary field $b$ would be 
\begin{equation}
     b ~=~ -~ 2 \tilde{\k} (\pa \l) \l \l \l 
     ~~~,
\end{equation}
which also takes the same form as the equation of motion of $F$ (ignoring terms with dynamical fields other than $\l$). Therefore treating $\pa A$ as an auxiliary field would not grant us the SYK term. This is inevitable as $F$ and $\pa A$ has the same engineering dimension to start with. 

The conclusion of this attempt is either we utilize the FI mechanism effectively via $\cl_{\rm FI-1}$, which would give us the SYK term and dynamical bosons, or we modify it to $\cl_{\rm FI-2}$ which would get rid of dynamical bosons but would not give us the SYK term.

\newpage
\section{Conclusion}
\label{sec:CONcL}

In \cite{GHM}, we constructed several 1D, $\cn=4$ supersymmetric SYK models based on chiral supermultiplet, vector supermultiplet, and tensor supermultiplet. In this work, we contribute to the discussion by including the complex linear supermultiplet. 

In addition to 3-point and $q$-point superfield vertices, we introduced a class of models based on the modification of the CLS kinetic terms in Section \ref{sec:QQ24pt}, which is allowed by introducing a copy of VS or TS through relaxing the complex linear constraint. This class of models has the special feature of containing the desired 4-point SYK terms off-shell (and they will stay when we go on-shell).

In section \ref{subsec:N2susy} we study the model with two 3-point superfield vertices, one with one anti-chiral and two chiral superfields, and another with the anti-chiral superfield changed to complex anti-linear superfield. When the coupling constants are chosen to be the same, the four-fermion terms vanish. This indicates a hidden $\cn=2$ supersymmetry in the model.

Another remark is on the incompatibility of SYK terms and the absence of dynamical bosons. The efforts described in Section \ref{sec:dynamicalbosons} indicate that 1D, ${\cal N}=4$ models with SYK-type interactions must include dynamical bosons. By removing dynamical bosons, we will also remove SYK terms.

\vspace{.05in}
 \begin{center}
\parbox{4in}{{\it ``My [algebraic] methods are really methods of working 
  \\ $~~$
and thinking; this is why they have crept in everywhere 
  \\ $~~$
 anonymously.''\\ ${~}$
\\ ${~}$ }\,\,-\,\, Amalie Emmy Noether }
 \parbox{4in}{
 $~~$} 
 \end{center}
 \noindent
{\bf {Acknowledgements}}\\[.1in] \indent

The research of S.\ J.\ G., Y.\ Hu, and S.-N.\ Mak is supported 
in part by the endowment of the Ford Foundation Professorship of Physics at Brown 
University and they gratefully acknowledge the support of the Brown Theoretical Physics 
Center. 
Work by S.-N.\ Mak is partially supported by Galkin Foundation Fellowship, and 
work by Y.\ Hu is partially supported by Physics Dissertation Fellowship at Brown University.

\newpage
\appendix

\section{Superspace Conventions\label{appen:convention}}
In this appendix, we review the two-component notation in 4D, ${\cal N} = 1$ superspace. First, the algebra that super-covariant derivatives satisfy is
\begin{align}
     \{ {\rm D}_{\a},\, {\rm D}_{\b} \} ~=&~ \{ \overline{{\rm D}}_{\dot \a},\, \overline{{\rm D}}_{\dot \b} \} ~=~ 0 \\
     \{ {\rm D}_{\a},\, \overline{{\rm D}}_{\dot \b} \} ~=&~ i\pa_{\a\dot\b}
\end{align}
where $\a,~\dot \a = 1,2$. 

\begin{align}
    {\rm D}^2 ~=&~ \fracm12 {\rm D}^{\a}\,{\rm D}_{\a} \\
    \overline{{\rm D}}^2 ~=&~ \fracm12 \overline{{\rm D}}^{\dot \a}\,\overline{{\rm D}}_{\dot \a} \\
    \Box ~=&~ \fracm12 \pa^{\a\dot\a}\,\pa_{\a\dot\a} \\
    \int d^2\theta d^2\overline{\theta} ~=&~ {\rm D}^2\,\overline{{\rm D}}^2 \,|_{\theta\to 0, \overline{\theta}\to 0}
\end{align}

\begin{align}
    ({\rm D}_{\a})^* ~=&~ -\overline{{\rm D}}_{\dot \a} \\
    ({\rm D}^{\a})^* ~=&~ \overline{{\rm D}}^{\dot \a} \\
    \rightarrow ({\rm D}^2)^* ~=&~ \overline{{\rm D}}^2
\end{align}
where $*$ means hermitian conjugation. 

Useful identities:
\begin{align}
    &{\rm D}_{\a}\,{\rm D}^2 ~=~ \overline{{\rm D}}_{\dot \a}\,\overline{{\rm D}}^2~=~ 0\\
  &{\rm D}^{\a}\,\overline{{\rm D}}^2~=~ \overline{{\rm D}}^2\,{\rm D}^{\a} ~+~ i\pa^{\a\dot\a}\,\overline{{\rm D}}_{\dot \a}\\
  &\overline{{\rm D}}^{\dot\a}\,{\rm D}^2~=~ {\rm D}^2\,\overline{{\rm D}}^{\dot\a} ~+~ i\pa^{\a\dot\a}\,{\rm D}_{ \a}\\
  & {\rm D}_{\a}\, {\rm D}_{\b} ~=~ -C_{\a\b}\, {\rm D}^2\\
  & {\rm D}^{\g}\, \overline{{\rm D}}^2\, {\rm D}_{\g} ~=~ \overline{{\rm D}}^{\dot \g}\, {\rm D}^2\, \overline{{\rm D}}_{\dot \g} ~=~ 2\,{\rm D}^2\,\overline{{\rm D}}^2 ~+~i\pa^{\a\dot\a}\,{\rm D}_{\a}\, \overline{{\rm D}}_{\dot \a}  \\
  &\overline{{\rm D}}^2\, {\rm D}^2~=~ {\rm D}^2\,\overline{{\rm D}}^2 ~+~ \Box ~+~ i\pa^{\a\dot\a}\,{\rm D}_{\a}\, \overline{{\rm D}}_{\dot \a} \\
  & {\rm D}^2\, \overline{{\rm D}}^2~=~ \overline{{\rm D}}^2\, {\rm D}^2 ~+~ \Box ~+~ i\pa^{\a\dot\a}\,\overline{{\rm D}}_{\dot \a} \,{\rm D}_{\a} 
\end{align}

We can relate a vector label $\underaccent{\tilde}{a}$ in a Cartesian coordinate basis to the $\a\dot\a$ basis by a set of Clebsch-Gordan coefficients, the Pauli matrices: 
\begin{align}
    {\rm For~ fields: }~ V^{\a\dot\a} ~=&~ \frac{1}{\sqrt{2}}\,(\s_{\underaccent{\tilde}{b}})^{\a\dot\a}\,V^{\underaccent{\tilde}{b}}~~,~~
    V^{\underaccent{\tilde}{b}} ~=~ \frac{1}{\sqrt{2}}\,(\s^{\underaccent{\tilde}{b}})_{\a\dot\a}\,V^{\a\dot\a}\\
    {\rm For~ derivatives: }~ \pa_{\a\dot\a} ~=&~ (\s^{\underaccent{\tilde}{b}})_{\a\dot\a}\,\pa_{\underaccent{\tilde}{b}}~~,~~
    \pa_{\underaccent{\tilde}{b}} ~=~ \frac{1}{2}\,(\s_{\underaccent{\tilde}{b}})^{\a\dot\a}\,\pa_{\a\dot\a}\\
    {\rm For~ coordinates: }~ x^{\a\dot\a} ~=&~\frac{1}{2}\, (\s_{\underaccent{\tilde}{b}})^{\a\dot\a}\,x^{\underaccent{\tilde}{b}}~~,~~
    x^{\underaccent{\tilde}{b}} ~=~ (\s^{\underaccent{\tilde}{b}})_{\a\dot\a}\,x^{\a\dot\a}
\end{align}

Pauli matrices satisfy:
\begin{align}
    (\s_{\underaccent{\tilde}{a}})^{\a\dot\a}\,(\s^{\underaccent{\tilde}{b}})_{\a\dot\a} ~=&~ 2\,\d_{\underaccent{\tilde}{a}}{}^{\underaccent{\tilde}{b}}\\
    (\s^{\underaccent{\tilde}{b}})^{\a\dot\a}\,(\s_{\underaccent{\tilde}{b}})_{\b\dot\b} ~=&~ 2\,\d_{\a}{}^{\b}\,\d_{\dot\a}{}^{\dot\b}
\end{align}

\newpage
$$~~$$


\begin{thebibliography}{99}
\small\frenchspacing\raggedright



\bibitem{GHM}
S.~J.~Gates, Y.~Hu and S.~N.~H.~Mak,
``On 1D, N = 4 Supersymmetric SYK-Type Models (I),''
[arXiv:2103.11899 [hep-th]].



\bibitem{SY}
S.~Sachdev and J.~Ye, 
``Gapless spin fluid ground state in a random, quantum
Heisenberg magnet,'' 
Phys. Rev. Lett. \textbf{70}, no. 21, 3339 (1993) 
doi:10.1103/PhysRevLett.70.3339
[arXiv:9212030 [cond-mat]].

\bibitem{K1}
A. Kitaev, 
``Hidden Correlations in the Hawking radiation and Thermal Noise,'' 
{\it KITP Fundamental Physics Prize Symposium} (Nov. 10, 2014),
http://oneline.kitp.ucsb.edu/online/joint98/.

\bibitem{K2}
A. Kitaev, 
``A simple model of quantum holography," 
{\it KITP strings seminar and
Entanglement 2015 program} (Feb. 12, Apr. 7, \& May 27, 2015),
http://oneline.kitp.ucsb.edu/online/entangled15/.



\bibitem{GL1}
Yu.\ A.\ Gol'fand and E. P. Likhtman, 
"Extension of the Algebra of Poincare Group Generators and violation of P Invariance,"
JETP Lett.\ {\bf {13}}
(1971) 323.

\bibitem{GL2}
Yu.\ A.\ Gol'fand and E. P. Likhtman, Pisma Zh.Eksp. Teor. Fiz {\bf {13}}
(1971) 452.

\bibitem{WZ}
J.\ Wess, and B.\ Zumino, ``A Lagrangian Model Invariant Under Supergauge 
Transformations,'' Phys.\ Lett.\ {\bf {B49}} (1974) 52; idem.\ ``Supergauge 
Transformations in Four-Dimensions,'' Nucl.\ Phys.\ {\bf {B70}} (1974) 39.

\bibitem{Wss}
J.\ Wess, Lectures given at the Bonn Summer School 1974.

\bibitem{Fy8}
P.\ Fayet, ``Fermi-Bose Hypersymmetry,'' Nucl.\ Phys.\ {\bf {B113}} (1976) 135.



\bibitem{VSM1}
J.\ Wess, and B.\ Zumino, Nucl.\ Phys.\ B78 (1974) 1;
DOI: 10.1016/0550-3213(74)90112-6.

\bibitem{VSM2}
S.\ Ferrara, and B.\ Zumino Nucl.\ Phys.\ 679 (1974) 413,
DOI: 10.1016/0550-3213(74)90559-8.



\bibitem{C-GuLL}
W.\ Siegel,
``Gauge Spinor Superfield as a Scalar Multiplet;''
 Phys.\ Lett.\ B 85 (1979) 333,
 DOI: 10.1016/0370-2693(79)91265-6. 




\bibitem{Neqv}
T.\ H\" ubsch, ``Linear and chiral superfields are usefully inequivalent,''
Class.\ Quant.\ Grav.\ {\bf {16}} (1999) L51-L54, DOI: 10.1088/0264-9381/16/9/101,
e-Print: hep-th/9903175.

\bibitem{TwsT}
  S.\  J.\ Gates, Jr., 
``Superspace Formulation of New Nonlinear Sigma Models,''
Nucl.\ Phys.\ B 238 (1984) 349,
 DOI: 10.1016/0550-3213(84)90456-5.

\bibitem{ThRR}
W. Thirring, 
{\bf {Ann.\ Phys.\ 3 \, (N.Y.)}} (1958),
``A Soluble relativistic field theory?,''
91, DOI: 10.1016/0003-4916(58)90015-0.

\bibitem{NmBJL1}
Y.\ Nambu, and G.\ Jona-Lasinio, 
``Dynamical Model of Elementary Particles Based on an Analogy with Superconductivity. I,''
{\bf {Phys.\ Rev.\ 122}} (1961) 345,  DOI: 10.1103/PhysRev.122.345.
 
\bibitem{NmBJL2}
Y.\ Nambu, and G.\ Jona-Lasinio, 
``Dynamical Model of Elementary Particles Based on an Analogy with Superconductivity. II,''
{\bf {Phys.\ Rev.\ 124}} (1961) 246,  DOI: 10.1103/PhysRev.124.246.
 
\bibitem{GN}
D.\ Gross, and A.\ Neveu, 
``Dynamical symmetry breaking in asymptotically free field theories,'' 
{\bf {Phys.\ Rev.\ D.\ 10 (10)}} (1974) 3235, 
DOI: 10.1103/PhysRevD.10.3235.




\bibitem{SFSG}
W.\ Siegel, and S.\ J.\ Gates, Jr., ``Superfield Supergravity,''
Nucl.\ Phys.\  {\bf {B147}} (1979) 77-104, DOI: 10.1016/0550-3213(79)90416-4. 

\bibitem{GS}
S.\ J.\ Gates, Jr., and W.\ Siegel,``Variant Superfield Representations,''
Nucl.\ Phys.\ {\bf {B187}} (1981) 389-396,
DOI: 10.1016/0550-3213(81)90281-9. 

\bibitem{SUSYBk}
S.\ J.\ Gates, Jr., M.\ T.\ Grisaru, M.\ R\v ocek, and W.\ Siegel.
{\em {Superspace Or One Thousand and One Lessons in 
Supersymmetry}}, Front.\ Phys.\ {\bf {58}} (1983) [hep-th/0108200].

\bibitem{CNM}
B.\ B.\ Deo and S.\ J.\ Gates Jr., ``Comments On Nonminimal N=1 Scalar Multiplets,''
Nucl.\ Phys.\ {\bf {B254}} (1985) 187.

\bibitem{B&K} 
I.\ L.\ Buchbinder, and S.\ M.\ Kuzenko,  
{\em {Ideas and methods of supersymmetry and supergravity: 
Or a walk through superspace}}. (Studies in High Energy 
Physics, Cosmology and Gravitation). CRC Press.
(1998), ISBN-13: 978-0750305068.

\bibitem{K1ps}
S.\ M.\ Kuzenko,
``Projective superspace as a double-punctured harmonic superspace,'' Int.\ J.\ Mod.\ Phys.\ A {\bf 14}, 1737 (1999), 
DOI: 10.1142/S0217751X99000889,
[arXiv:hep-th/9806147].

\bibitem{CNM1}
S.\ J.\ Gates, Jr. and S.\ M.\ Kuzenko, ``The CNM Hypermultiplet Nexus,'' Nucl.\  Phys.\ 
{\bf {B543}} (1999) 122, [arXiv:hep-th/9810137].

\bibitem{CNM2}
S.\ J.\ Gates, Jr., T.\ H\"ubsch, S.\ M.\  Kuzenko, ``CNM Models, Holomorphic Functions and Projective Superspace C Maps,''
Nucl.Phys. {\bf {B557}} (1999) 443, [arXiv:hep-th/9902211]; 

\bibitem{CNM3}
S.\ J.\ Gates, Jr. and S.\ 
M.\ Kuzenko, ``4-D, N=2 Supersymmetric Off-Shell Sigma Models on the Cotangent 
Bundles of K\" ahler Manifolds,'' 
Fortsch.\ Phys.\  {\bf 48}, 115 (2000), DOI: 10.1002/(SICI)1521-3978(20001)48:1/3<115::AID-PROP115>3.0.CO;2-F",
[arXiv:hep-th/9903013].

\bibitem{K4}
S.\ J.\ Gates Jr., T.\ H\" ubsch, and S.\ M.\ Kuzenko, ``CNM models, holomorphic functions and projective superspace C maps
Nucl.\ Phys.\ B 557 (1999) 443, DOI: 10.1016/S0550-3213(99)00370-3,
[arXiv:hep-th/9902211].

\bibitem{CNM4}
S.\ J.\ Gates, Jr., S.\ Penati, and G.\ Tartaglino-Mazzucchelli, ``6D Supersymmetry, Projective 
Superspace \& 4D $\cal N$ = 1 Superfields,'' JHEP 0609 (2006) 006 e-Print: hep-th/0604042 
[hep-th]; idem., ``6D supersymmetry, projective superspace and 4D, N = 1 superfields,'' 
JHEP 0605 (2006) 051, e-Print: hep-th/0508187 [hep-th].

\bibitem{CNM4a}
S.\ J.\ Gates, Jr., S.\ Penati, and G.\ Tartaglino-Mazzucchelli, ``6D Supersymmetric Nonlinear Sigma-Models in 4D, N=1 Superspace,''
JHEP 09 (2006) 006, DOI: 10.1088/1126-6708/2006/09/006,
 arXiv:: hep-th/0604042 [hep-th]

\bibitem{K5}
M.\ Arai and M.\ Nitta, ``Hyper-K\"ahler sigma models on (co)tangent bundles with SO(n) isometry,''
Nucl.\ Phys.\ B {\bf 745}, 208 (2006), DOI: 10.1016/j.nuclphysb.2006.03.033,
[arXiv:hep-th/0602277].

\bibitem{K6}
M.\ Arai, S.\ M.\ Kuzenko and U.\ Lindstr\"om, ``Hyperk\"ahler sigma models on cotangent bundles of Hermitian symmetric
spaces using projective superspace,''
JHEP {\bf 0702}, 100 (2007), DOI: 10.1063/1.2823784,
[arXiv:hep-th/0612174].

\bibitem{K7}
M.\ Arai, S.\ M.\ Kuzenko and U.\ Lindstr\"om,  ``Polar supermultiplets, Hermitian symmetric spaces and hyperk\"ahler
metrics,''  JHEP {\bf 0712}, 008 (2007), DOI: 10.1063/1.2823784,
[arXiv:0709.2633 [hep-th]].

\bibitem{K8}
S.\ M.\ Kuzenko and J.\ Novak, ``Chiral formulation for hyperk\"ahler sigma-models on cotangent bundles of
symmetric spaces,''  JHEP {\bf 0812}, 072 (2008), DOI: 10.1088/1126-6708/2008/12/072,
[arXiv:0811.0218 [hep-th]].

\bibitem{K9}
S.\ M.\ Kuzenko,  ``Lectures on nonlinear sigma-models in projective superspace,''
J.\ Phys.\ A {\bf A43}, 443001 (2010), DOI: 
10.1088/1751-8113/43/44/443001,
[arXiv:1004.0880 [hep-th]].

\bibitem{Adhm}
S.\ James Gates Jr., and Lubna Rana,
``Manifest (4,0) supersymmetry, sigma models and the ADHM instanton construction,"
Phys.\ Lett.\ {\bf {B345}} (1995) 233, DOI: 10.1016/0370-2693(94)01653-T,
[arXiv: 9411091 [hep-th]].


\bibitem{hetN4b}
R.\ Dhanawittayapol, S.\ James Gates Jr., Lubna Rana,
``A Canticle on (4,0) Supergravity-Scalar Multiplet Systems for a ``Cognoscente''
Phys.\ Lett.\ {\bf {B389}} (1996) 264, DOI:  0.1016/S0370-2693(96)01254-3,
[arXiv: 9606108 [hep-th]].


\end{thebibliography}
\end{document}